\documentclass[12pt, draftclsnofoot, onecolumn]{IEEEtran}
\usepackage[font=small,skip=0.5pt]{caption}
\usepackage{amsmath,amssymb}
\usepackage{amsfonts}
\usepackage{amsbsy}
\usepackage{graphicx}

\usepackage{subfigure}
\usepackage{stfloats}
\usepackage{epstopdf}
\usepackage{color}
\usepackage{etoolbox}
\usepackage{upgreek}

\makeatletter
\patchcmd{\@maketitle}
{\addvspace{0.5\baselineskip}\egroup}
{\addvspace{-1\baselineskip}\egroup}
{}
{}
\makeatother
\begin{document}
	
	\title{\huge Overlay Space-Air-Ground Integrated Networks with SWIPT-Empowered Aerial Communications}
	\author{Anuradha~Verma, Pankaj~K.~Sharma, Pawan~Kumar, and~Dong~In~Kim
		\thanks{Anuradha Verma, Pankaj K. Sharma, and Pawan Kumar are with the Department
			of Electronics and Communication Engineering, National Institute of
			Technology Rourkela, Rourkela 769008. Email: \{519ec1026, sharmap, kumarpa\}@nitrkl.ac.in.}
		\thanks{Dong In Kim is with the Department of Electrical and Computer Engineering, Sungkyunkwan University, Suwon 16419, South Korea. Email: dikim@skku.ac.kr.
		}
	}
	
	\maketitle
	
	\begin{abstract}
	In this article, we consider overlay space-air-ground integrated networks (OSAGINs) where a low earth orbit (LEO) satellite communicates with ground users (GUs) with the assistance of an energy-constrained coexisting air-to-air (A2A) network. Particularly, a non-linear energy harvester with a	hybrid SWIPT utilizing both power-splitting and time-switching energy harvesting (EH) techniques is employed at the aerial transmitter. Specifically, we take the random locations of the satellite, ground and aerial receivers to investigate the outage performance of both the satellite-to-ground and aerial networks leveraging the stochastic tools. By taking into account the Shadowed-Rician fading for satellite link, the Nakagami-\emph{m} for ground link, and the Rician fading for aerial link, we derive analytical expressions for the outage probability of these networks. For a comprehensive analysis of aerial network, we consider both the perfect and imperfect successive interference	cancellation (SIC) scenarios. Through our analysis, we illustrate that, unlike linear EH, the implementation of non-linear EH	provides accurate figures for any target rate, underscoring the	significance of using non-linear EH models. Additionally, the influence of key parameters is emphasized, providing guidelines	for the practical design of an energy-efficient as well as spectrum-efficient future non-terrestrial networks. Monte Carlo simulations validate the accuracy of our theoretical developments.

\end{abstract}
	
	\section{Introduction}
	\IEEEPARstart{T}{he} fifth-generation (5G) wireless networks are being deployed worldwide to enable vast applications and services that have not been enabled by the previous generations. However,  5G networks would face significant challenges in the future due to excessive growth in mobile traffic and the emergence of advanced use cases, e.g., holographic communications, tactile internet, intelligent autonomous transportation, digital twin, global ubiquitous connectivity, etc. \cite{1}. Therefore,  researchers from both academia and industry have recently directed their attention towards the development of beyond 5G (a.k.a. sixth-generation (6G)) networks \cite{2}, \cite{3}. The 6G networks aim to support advanced applications under three major classes \cite{4}, viz., ubiquitous mobile ultra-broadband (uMUB),  ultra-high data density (uHDD), and ultra-high speed with low latency communications (uHSLLC) that require peak data rates 10 to 100 times larger and the latency 10 times lesser than the 5G. The important 6G targets, i.e., global ubiquitous connectivity, extreme ultra reliability, energy efficiency, etc. cannot be effectuated by the 5G ground-based networks (GNs) alone. A non-terrestrial network (NTN) architecture embracing the low earth orbit (LEO) satellites integrated with ground-based and/or aerial networks have been envisaged as a key enabler for 6G \cite{6}-\cite{5}. Space-ground integrated network (SGIN) architecture is a common variant of NTNs that targets simple integration of space and ground network layers for satellite-to-ground (S2G) communications. Several commercial projects on SGINs, e.g., Starlink, Global Information Grid, OneWeb, Telesat, etc., have already been undertaken by the industry \cite{5}. In general, the direct satellite links in SGINs experience large attenuation due to clouds, rain, fog, physical blockage, indoor users, etc. Therefore, the SGINs make use of cooperative relaying in the ground network layer for reliable communications. The coverage provided by a ground-based cooperative node is rather limited due to restricted deployment under certain economic and geographic constraints. Thus, the SGIN architecture is advanced by inducting an aerial network layer between space and ground network layers comprising high altitude platforms, balloons, and/or unmanned aerial vehicles (UAVs). This generalised NTN architecture is widely recognized as space-air-ground integrated networks (SAGINs) \cite{7}-\cite{11}, where the aerial nodes assist S2G communications by acting as a cooperative relay when the direct satellite link is severely attenuated. Recently, UAVs have been qualified as potential candidates for the role of aerial relays due to their agility, line-of-sight (LoS) connectivity, and flexible deployment. Despite the advantages, the service time of UAVs is curbed by the limited size of their onboard battery \cite{14}.  Simultaneous wireless information and power transfer (SWIPT) \cite{15}, \cite{18} is an off-the-shelf technology that may be utilised for partially supplying energy to UAV's onboard communication and sensor circuits to prolong its service time. However, the current satellite technology is not yet mature for reliable wireless power transfer (WPT) in SAGINs. The major roadblock here is the large distance between satellite and ground users (GUs). Nevertheless, the development of more powerful LEO satellites equipped with multi-antenna in the future is expected to feature WPT functionality for SAGINs. In \cite{15}, \cite{16}, the authors have already advocated to explore WPT techniques for S(A)GINs. 
	
	Besides, the aerial network layer of SAGINs supports air-to-air (A2A) networks where UAVs are deployed as aerial users \cite{19}. It is foreseeable that the number of aerial users would increase dramatically in a few years with UAVs finding their way to internet-of-things \cite{22}. This may put the spectrum resources at the risk of shortage. The spectrum sharing in SAGINs can essentially resolve this problem, where a primary (licensed) S2G network and a secondary (unlicensed) A2A network coexist in the same spectrum resources \cite{20}. The overlay spectrum sharing model \cite{23}, \cite{24} can be of great significance for SAGINs, where an A2A network can be accommodated within the licensed spectrum of satellite network in exchange for assisting S2G communications. In order to tap the combined potential of WPT and spectrum sharing techniques for SAGINs, an OSAGIN architecture with SWIPT-empowered integrated A2A network is of our specific interest in this article. It is worth mentioning that the OSAGINs can be implemented in a simple dual-hop cooperative spectrum sharing framework with less stringent transmit power constraints as compared to its underlay counterpart. 

	\subsection{Prior Arts}		
	The SGINs have been extensively studied in recent years \cite{35}-\cite{39}. For instance, the authors in \cite{35} have investigated the performance of an SGIN by employing an amplify-and-forward (AF) relay. In \cite{36}, the OP and throughput of the multi-antenna multi-relay SGINs have been analysed.  The work in \cite{37} has evaluated the performance of a downlink SGIN with non-orthogonal multiple access (NOMA). In \cite{38}, the authors have considered a multiuser SGIN to examine its OP and ergodic capacity. The work in \cite{39} has analysed the OP of an SGIN in the presence of multiple two-way relays. Further, the works in \cite{40}-\cite{42} have capitalized on the LEO satellites for SGINs to achieve larger coverage, lower transmission losses, and higher bandwidth. There are a handful of research works \cite{21}-\cite{45} that have incorporated spectrum sharing in SGINs for enhanced spectral efficiency. The authors in \cite{21} have highlighted the importance of spectrum sharing in 6G SGINs based on NOMA.   Reference \cite{44} has explored secure transmission within a cognitive SGIN featuring a multi-antenna eavesdropper. More importantly, the authors in \cite{23}, \cite{24}, \cite{43}, \cite{45} have relied on the overlay spectrum sharing for SGINs. Further, the WPT in SGINs has been considered in \cite{54}. For spectrum sharing SGINs, the WPT has been considered by the authors in \cite{17}-\cite{47}. So far, all these works have focused on the SGINs without taking into account the aerial relays/users. The SAGINs incorporating the aerial relays have been investigated in \cite{28}-\cite{50}. The works in \cite{28}-\cite{27} have analysed the performance of SAGINs with a single aerial relay. The SAGINs with multiple aerial relays have been considered in \cite{25} and \cite{29}. In \cite{30}, an innovative SAGIN has been proposed with spatially random aerial/ground terminals lying within the cone-shaped satellite beam. The work in \cite{31} has adopted a stochastic geometry-based framework for SAGINs with downlink, uplink, and inter-aerial relaying links. In \cite{34}, a unified approach for the performance analysis of SAGINs was presented. Further, in \cite{48}, a spectrum sharing-based SAGIN has been analysed in the presence of imperfect channel state information. On the contrary, the work in \cite{50} has addressed a sum-rate maximization problem for SAGINs, where UAV-assisted WPT has been considered for enabling uplink transmission from GUs to satellite. Note that the aforementioned works have ignored the WPT for aerial relays/nodes in SAGINs to improve their service time. As argued previously, the spectral-efficient SAGINs are crucial for the coexistence of futuristic A2A networks within the fixed spectrum resources. In addition, the WPT is vital for extending the service time of energy-constrained A2A networks. Recently, we have proposed an OSAGIN in \cite{53}, where a primary S2G network and a SWIPT-empowered secondary A2A network are integrated to facilitate concurrent S2G and A2A communications. Herein, the S2G communication from a LEO satellite randomly located onto a three-dimensional (3D) spherical surface around the Earth to a GU is assisted by an A2A network that comprises an aerial transmitter (ATx) and aerial receiver (ARx) pair. In the A2A network, the ATx is assumed to be SWIPT-empowered that first harvests energy from the satellite’s signal and then, utilises it to broadcast a network-coded signal (containing both the satellite's and ATx's signals) for the intended GU and ARx. However, in this preliminary work, we have considered an idealistic linear energy harvester at the ATx for realizing SWIPT. Also, it assumes a simplified two-dimensional (2D) morphology for the A2A network which accounts for the randomness of ARx positions in a 2D disc. It is worth pointing out that a practical energy harvesting circuit always bears non-linear characteristics \cite{51}, \cite{52} and thus, a linear energy harvester disqualifies for an accurate estimation of the harvested power. Further, the ATx of an A2A network basically serves an ARx lying inside a 3D region illuminated by its electromagnetic beam toward the ground plane. In fact, a beam-inspired 3D configuration for an A2A network is more general, especially when the ARx has 3D mobility.  
	
	\subsection{Contributions}
	In this paper, we extend our preliminary work \cite{53} to a more realistic and generalized set-up by considering a non-linear energy harvesting circuit and a beam-inspired 3D configuration for the A2A network. The main contributions of the paper can be summarized as follows:
	\begin{itemize}
		\item We characterise the statistical distribution of distances from a LEO satellite to ATx, ATx to ARx, and ATx to GUs to represent the path loss corresponding to these links in a realistic manner. Specifically, we obtain the statistical distribution of the distance between ATx and an LEO satellite uniformly located onto a 3D surface around the Earth. We obtain the statistical distribution of distance between ATx and the ARx uniformly located inside a cone-shaped 3D region illuminated by the ATx's electromagnetic beam toward the ground plane. We finally obtain the distribution of distance between ATx and the GUs uniformly located inside the circular beamspot of ATx on the ground plane.

		\item We consider that the ATx is SWIPT-empowered and equipped with a non-linear energy harvesting circuit. We employ a hybrid energy harvesting protocol (to be discussed later) that simultaneously exploits the time-switching (TS) and power-splitting (PS) techniques to assist S2G transmissions while realizing the A2A communications. Here, the ATx relies on a simple amplify-and-forward (AF) protocol to retransmit the satellite signal combined with its own message signal. 
		
		\item By considering shadowed-Rician (SR), Nakagami-\emph{m}, and Rician fading pertaining to the space, air-to-ground, and A2A links, respectively, we evaluate the OP of both the S2G and A2A networks of the proposed OSAGIN. We hereby carry-out the OP analysis by taking into account both the imperfect interference cancellation (im-IC) and the perfect interference cancellation (p-IC) mechanisms at the ARx. We draw useful insights based on our numerical results for the considered system model.
		
	\end{itemize}
	\subsection{Paper Organization and Useful Notations}
	The rest of the article is organized as follows: Section II describes the system and channel models, relevant distance distributions, and hybrid energy harvesting protocol. Section III presents the OP analysis of the S2G network. Sections IV contain the OP analysis of the A2A network under im-IC and p-IC. Section V discusses  the numerical result and draw useful insights. Finally, Section VI draws the conclusions. Table \ref{table:Table1} provides the summary of notations used in this paper. 
	\begin{table}[!t]
		\centering
		\caption{Summary of notations used in the paper.}
		\label{table:Table1}
		\begin{tabular}{c||c}
			\hline\hline
			\textbf{Notation} & \textbf{Definition}  \\ \hline\hline
			$\chi$    &  Energy conversion efficiency \\ \hline
			$\mathcal{P}_\textmd{th}$	 &Saturation threshold
			power of the EH circuit \\ \hline
			$\rho$ & Time splitting factor  \\ \hline
			$\varepsilon$ & Power splitting factor\\ \hline
			
			$\mathcal{K}_{ \upsilon}(\cdot)$ & Modified Bessel function of second kind 
			\\ \hline	Pr$(\cdot )$ &  Probability \\ \hline
			$F_{X}(\cdot )$ and  $f_{x}(\cdot )$ & CDF and PDF\\ \hline
			$\Gamma (\cdot, \cdot)$ & Upper incomplete Gamma function \\ \hline
			$\mathbb{W}_{a,b}(\cdot )$ & Whittaker function \\ \hline
			$I_0(\cdot)$ & zeroth-order modified Bessel function of first kind \\ \hline
			$G^{m,n}_{p,q}\left[\cdot\vert\cdot\right]$ & Meijer G-function \\ \hline
			$\mu $   & Spectrum sharing factor \\ \hline
			$\chi_{\rho,\varepsilon}$ & $\chi( {2\rho }/{(1-\rho) } +\varepsilon )$\\ \hline
			$\mu^\prime$ & $\mu/(1-\mu)$ \\ \hline
			$\bar{\beta}_{sr}$&$\beta_{sr} -\delta_{sr}$ \\ \hline
			$\mathcal{A}$&$\chi_{\rho,\varepsilon}\eta_s(\mu-(1-\mu)\gamma_{\textmd{S}})$\\ \hline
			$\mathcal{B}$& $\mu_\varepsilon\chi_{\rho,\varepsilon}\gamma_{\textmd{S}}$, \\ \hline
			$\mathcal{C}$&$\chi_{\rho,\varepsilon}\eta_s((1-\mu)-\mu\gamma_{\textmd{A}})$\\ \hline
			$\mathcal{D}$&$\mu_\varepsilon\chi_{\rho,\varepsilon}\gamma_\textmd{A}$ \\ \hline $\mathcal{E}$&$\chi_{\rho,\epsilon}\eta_s(1-\mu)$\\ \hline
			$\mathcal{P}_{\mathcal{A}, \mathcal{B}}$& ${\mathcal{P}_\text{th}\mathcal{A}}/{\eta_{s}}-\mathcal{B}$\\ \hline
			$ \mathcal{P}_{\mathcal{C}, \mathcal{D}}$&${\mathcal{P}_\text{th}\mathcal{C}}/{\eta_{s}}-\mathcal{D}$ \\ \hline 
			$\Delta_\Gamma(a,b,c)$& $\Gamma(a,b)-\Gamma(a,c)$ \\ \hline 
			$\varsigma_1$&$1+k_1+k_2-n$\\ \hline 
			$\varsigma_2$&${(n
				+k_1+2k_2+1)}/{2}$\\ \hline
			$\varsigma_3$&${(-n+k_1+1)}/{2}$\\ \hline 
		\end{tabular} 
	\end{table}
	\section{System Description}\label{sysmod}
	
	\subsection{System Model}
	
\begin{figure}[!t]
		\centering
		\includegraphics[width=2.8in]{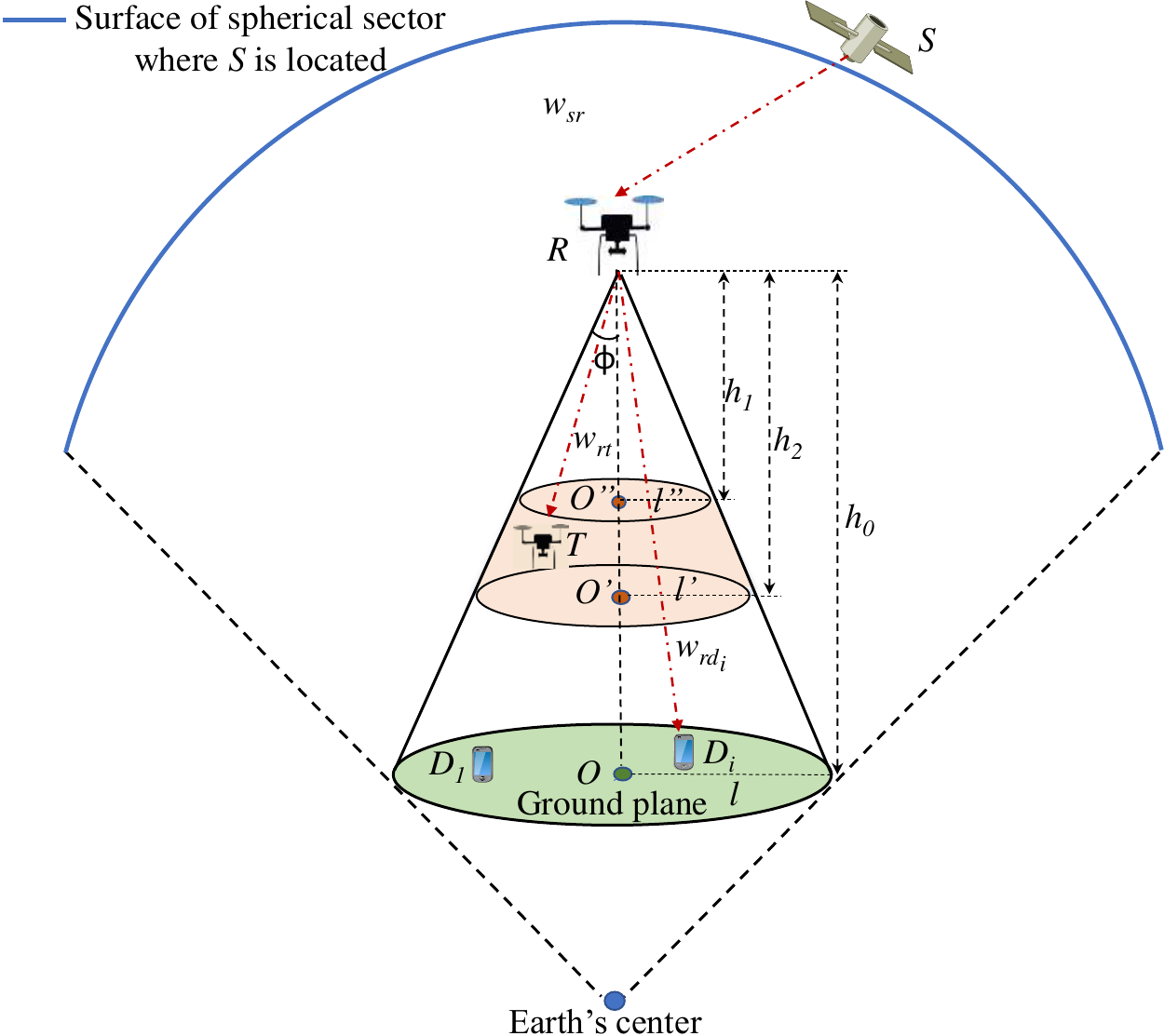}
		\caption{OSAGIN system model with SWIPT-enabled A2A communications.}
		\label{system}
	\end{figure}

	As shown in Fig. \ref{system}, we consider an OSAGIN that consists of a LEO satellite $S$, $N$ GUs $D_{i}$, $i \in \{1,...,N \}$, and a coexisting energy-constrained A2A network that comprises an ATx $R$ and an ARx $T$. We assume that the orbit of $S$ is the surface of a sphere\footnote{LEO-based satellite networks deploy multiple satellites in a grid pattern to cover a specific ground location. As a result, a serving LEO satellite is located within a 3D region rather than a fixed point in space.} from the centre of earth $E$ where $S$ is uniformly located. The GUs $D_{i}$, $i \in \{1,...,N \}$ are assumed to be clustered and uniformly located inside a disc $\mathbb{B}(O,l)$ onto the ground plane. We further assume that the ATx $R$ is statically hovering at an altitude of $h_0$ vertically above the centre of $\mathbb{B}(O,l)$. We consider that the ATx $R$ has a beam-inspired conical coverage underneath with half-beamwidth angle $\upphi$ that exactly overlaps $\mathbb{B}(O,l)$ onto the ground plane. Furthermore, the ARx $T$ is considered to be a mobile node uniformly located within the 3D conical coverage region below $R$ in between the discs $\mathbb{B}(O^{\prime\prime},l^{\prime\prime})$ and $\mathbb{B}(O^{\prime},l^{\prime})$ lying at distances $h_1$ and $h_2$, respectively. In the considered OSAGIN, the satellite $S$ sends its message signal to one of the GUs $D_i$ with the help of coexisting A2A network when the direct link between $S$ and $D_i$ is masked\footnote{Such situations arise under heavy clouds, rain, GUs located in indoor, etc.}. Here, we assume that the ATx $R$ is SWIPT-empowered and capable of overlay spectrum sharing\footnote{The nodes $S$ and $D_i$ together form a primary S2G network that uses licensed spectrum. However, the A2A network does not have access to any licensed spectrum for its operations. As a result, it depends on the S2G network's spectrum sharing for its own communication, while assisting the satellite communications.}, so that it can communicate with the ARx $T$ by simultaneously assisting the satellite transmissions. All nodes are assumed to be equipped with a single antenna and operate in the half-duplex mode. The euclidean distances pertaining to the links $S$ to $R$, $R$ to $D_i$, and $R$ to $T$ are denoted as $w_{sr}$, $w_{rd_i}$, and $w_{rt}$, respectively. Further, the channel coefficients for the links $S$ to $R$, $R$ to $D_i$, and $R$ to $T$ are denoted as $g_{sr}$, $g_{rd_i}$, and $g_{rt}$, respectively. We assume the block fading model for channels where they remain constant during a block and change independently in the next block. The thermal noise at each receiver node is assumed to be additive white Gaussian noise (AWGN).       
	
	
	\subsection{Channel Models}
	\subsubsection{Satellite-to-ATx link}
	The channel $g_{sr}$ follows SR fading and hence, the probability density function (pdf) of $|g_{sr}|^{2}$ can be given as
	\begin{align}\label{eq1}
		f_{|g_{sr}|^{2}}(x)=\alpha_{sr} \sum_{k=0}^{m_{sr}-1}\zeta(k)x^{k}\textmd{e}^{-(\beta_{sr}-\delta_{sr})x},
	\end{align}
	where $\alpha_{sr}=(2\flat_{sr} m_{sr}/(2\flat_{sr} m_{sr}+\Omega_{sr}))^{m_{sr}}/2\flat_{sr} $, $\beta_{sr}=1/2\flat_{sr} $, and $\delta_{sr}=\Omega_{sr}/(2\flat_{sr} )(2\flat_{sr} m_{sr}+\Omega_{sr})$, $\Omega_{sr}$ and $2\flat_{sr} $ are the average powers, respectively, of the LOS and multipath components, $m_{sr}$ is the fading severity,  $\zeta(k)=(-1)^{k}(1-m_{sr})_{k}\delta_{sr}^{k}/(k!)^{2}$, and $(\cdot)_{k}$ is Pochhammer symbol \cite[p. xliii]{grad}.
	
	Further, a scale factor corresponding to free space loss for the satellite link is given as \cite{24}, \cite{36} 
	\begin{align}\label{eq2}
		\sqrt{\mathrm{C}w_{sr}^{-2}\vartheta_s\vartheta(\theta_{sr})}=\frac{\xi\lambda\sqrt{\vartheta_s\vartheta(\theta_{sr})}}{4\pi w_{sr}\sqrt{{\mathcal{K_{B}}\mathbb{T}\mathcal{W}}}},
	\end{align}
	where $\mathcal{K_B}=1.38\times10^{-23}$J/K is the Boltzmann constant, $\mathbb{T}$ is the receiver noise temperature, $\mathcal{W}$ is the carrier bandwidth, $\xi$ is the rain attenuation coefficient, $\lambda$ is the carrier wavelength, and $w_{sr}$ is the same as defined previously. Here, $\vartheta_s$ denotes the antenna gain at satellite, $\vartheta(\theta_{sr})$ gives the beam gain of satellite towards $R$ which can be expressed as
	\begin{align} \label{eq3}
		\vartheta(\theta_{sr})=\vartheta_{sr}\left(\frac{\mathcal{J}_1(\rho_{sr})}{2\rho_{sr}}+36\frac{{J}_3(\rho_{sr})}{\rho^3_{sr}}\right),
	\end{align} 
	where $\theta_{sr}$ is the angular separation of $R$ from the satellite beam center, $\vartheta_{sr}$ is the antenna gain at $R$, $\mathcal{J}_\varrho(\cdot)$, $\varrho\in\{1,3\}$ is the Bessel function, and $\rho_{sr}=2.07123\frac{\sin \theta_{sr}}{\sin \theta_{{sr}3\textmd{dB}}}$ with $\theta_{{sr}3\text{dB}}$ as $3$dB beamwidth.
	
	\subsubsection{ATx-to-GU link}
	The Nakagami-\emph{m} fading is considered for the channel ${g}_{r{d_i}}$ which yields the pdf of $|{g}_{r{d_i}}|^2$ as 
	\begin{align}\label{eq4}
		f_{|g_{rd_i}|^2}(x)&=  \frac{{m_{rd_i}}^{m_{rd_i}}}{\Gamma(m_{rd_i})}\,x^{m_{rd_i}-1}\exp(-m_{rd_i}x)  ,
	\end{align}
	where $m_{rd_i}$ represents the fading severity parameter.
	
	Further, the path loss for the link can be given as $w^{-\nu_{rd_i}}_{rd_{i}}$, where $\nu_{rd_i}$ is the path loss exponent.  
	
	\subsubsection{ATx-to-ARx link}
	The Rician fading is considered for the channel ${g}_{r{t}}$ which results in the pdf of $|g_{rt}|^2$ as  
	\begin{align}\label{eq5}
		f_{|{g}_{r{t}}|^2}(x)&={(1+K_{rt})}\exp\left(-K_{rt} -{(1+K_{rt})x} \right)\mathcal{I}_0\left(2\sqrt{{K_{rt}(1+K_{rt})x}}\right),
	\end{align}
	where $K_{rt}$ is the Rician $K$-factor and $I_{0}(z)=\sum_{m=0}^{\infty }\frac{1}{(m!)^{2}}\left( \frac{z}{2} \right)^{2m}$ \cite[eq. 8.447]{grad} is the zeroth-order modified Bessel function of first kind \cite[eq. 9.6.10]{grad}. 
	
	Moreover, the path loss for the link can be given as $w^{-\nu_{rt}}_{rt}$, where $\nu_{rt}$ is the path loss exponent.  
	
	\subsection{ Relevant Distance Distributions}
\begin{figure}[!t]
		\centering
		\includegraphics[width=2.0 in]{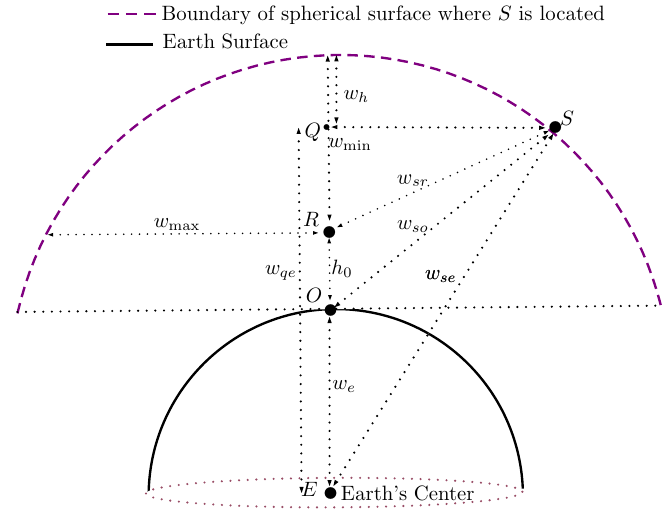}
		\caption{Three-dimensional geometry of the system model.}
		\label{leo}
	\end{figure}

	To enable OP analysis in the forthcoming sections, we provide a depiction of the geometry of the considered network being studied in Fig.~\ref{leo}. 
	\subsubsection{Distance Distribution for Satellite-to-ATx Link }
	We statistically characterize the distance between $S$ and $R$ in the following lemma. 
	\newtheorem{lemma}{Lemma}
	\begin{lemma}\label{lem}
		The probability density function (pdf) of the distance $w_{sr}$ is determined as
		\begin{align}\label{eq6a}
			f_{w_{sr}}(w)=\frac{w}{w_{er}w_{\min}},~ w_{\min}\leq w \leq w_{\max}, 
		\end{align}
		where  $w_{er}=w_{e}+w_{r}$ and $w_{\max}=\sqrt{w_{\min}^{2}+2{w_{er}w_{\min}}}$.  Hereby, $w_{\min}$ and $w_{\max}$ are the minimum and maximum distances between $S$ and $R$, and $w_e$ is the Earth's radius.
	\end{lemma}
	\begin{IEEEproof}
		Please refer to the Appendix A.
	\end{IEEEproof} 
	\subsubsection{Distance Distribution for ATx-to-GU Link }
	As followed in \cite{26}, the pdf of the distance between $R$ and $D_i$, i.e., $w_{rd_{i}}$ can be given as 	
	\begin{align}\label{eq7}
		f_{w_{rd_{i}}}(v) &=\frac{2v}{l^{2}},~h_0\leq v\leq\sqrt{h_0^{2}+l^{2}}.
	\end{align}
	\subsubsection{Distance Distribution for ATx-to-ARx Link }\label{ald}	
	It is worth mentioning that the ARx $T$ in the considered system model is assumed to be uniformly located inside a conic section formed by truncating the conical beam coverage of ATx $R$. As such, the pdf of uniformly distributed 3D locations for an aerial node inside a truncated cone was recently obtained in \cite{31} under the following two cases: 
	\newline\textbf{Case 1}: If $\frac{h_1}{\cos \upphi}<h_2$,
	\begin{equation}\label{eq8}
		{f_{{w_{rt}}}}({u }) =\left\{ \begin{array}{ll}
			\frac{6u({{u}-{h_{1}}})}{h_{2}^{3}-h_{1}^{3}}, &{\text{for }} {h_{1}} < u < \frac {{h_{1}}}{\cos \upphi }, \\
			\frac{6u^{2} ({1-\cos \upphi })}{h_{2}^{3}-h_{1}^{3}}, &{\text{for }}\frac {{h_{1}}}{\cos \upphi } < u < {h_{2}},\\ 
			\frac{6u({{h_{2}}-u\cos \upphi })}{h_{2}^{3}-h_{1}^{3}}, &{\text{for }}{h_{2}} < u< \frac {{h_{2}}}{\cos \upphi }.
		\end{array}\right.
	\end{equation}
	\newline\textbf{Case 2}: If $\frac{h_1}{\cos \upphi}>h_2$,
	\begin{equation}\label{eq8}
		{f_{{w_{rt}}}}({u }) =\left\{ \begin{array}{ll}
			\frac{6u({{u}-{h_{1}}})}{h_{2}^{3}-h_{1}^{3}}, &{\text{for }}{h_{1}} < u < h_2, \\
			\frac{6u({{h_2 }-{h_1 } })}{h_{2}^{3}-h_{1}^{3}}, &{\text{for }}{h_2} < u < \frac {{h_{1}}}{\cos \upphi },\\ 
			\frac{6u({{h_{2}}-{u}\cos \upphi })}{h_{2}^{3}-h_{1}^{3}}, &{\text{for }}~\frac {{h_{1}}}{\cos \upphi } < u < \frac {{h_{2}}}{\cos \upphi }.
		\end{array}\right.
	\end{equation}

	\subsection{Propagation Model with Hybrid Energy Harvesting}
	In the considered OSAGIN, the information transmission from satellite $S$ to GU $D_i$ takes place in two hops via ATx $R$. Let $R$ be an energy-constrained device whose onboard battery power is available only for the hovering purpose and thus, for communication, it relies on the energy harvesting from radio-frequency signals. Here, ATx $R$ adopts a hybrid protocol for energy harvesting and information processing to assist transmission from $S$ to $D_i$ while enabling the A2A transmission from $R$ to $T$. As such, the protocol is termed as hybrid because it exploits both the TS and PS techniques together for realizing the end-to-end communications. 
	\begin{figure}[!t]	
		\centering
		\includegraphics[width=3.4in]{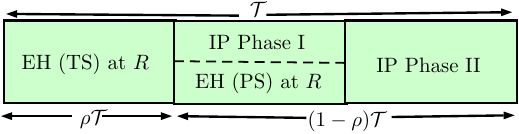}
		\caption{Illustration of the key parameters in the hybrid protocol for energy harvesting and information processing at $R$.}
		\label{bloc}
	\end{figure}
	
	The time-block structure for the adopted hybrid protocol at $R$ is illustrated in Fig. \ref{bloc}. Here, the entire transmission block of duration $\mathcal{T}$ is split into three consecutive time phases of durations $\rho\mathcal{T}$, $(1-\rho)\mathcal{T}/2$, and $(1-\rho)\mathcal{T}/2$, where $ \rho\in[0,1]$  represents the fraction of time for which the node $R$ harvests the energy. According to the hybrid protocol, in the first phase, the $S$ transmits its information signal $x_s$ to $R$ with power $P_s$ as given by
	\begin{align}\label{eq6}
		y_{r}&=\sqrt{P_{s}\vartheta_s\vartheta(\theta_{sr})\mathrm{C}w_{sr}^{-2}}g_{sr}x_{s}+n_{r},
	\end{align}
	where  $n_{r}\sim \mathcal{\mathbb{CN}}\left( 0,\sigma_{r}^{2} \right)$ represents the AWGN. In this phase, the $R$ harvests energy (without information processing) from $y_{r}$ based on the TS principle which is given by 
	\begin{align}\label{eq7}
		E^{TS}=\chi\rho \mathcal{T}\eta_s\left | g_{sr}\right |^{2}w_{sr}^{-2},\end{align}
	where, $\chi$ is the energy conversion efficiency with $\chi\in(0,1)$, and $\eta_s=P_{s}\mathrm{C}\vartheta_s\vartheta(\theta_{sr})$. 
	
	In the second phase (designated as IP Phase I), the $R$ again harvests a fraction of energy from $y_r$ (i.e., $\left | \sqrt{\varepsilon} y_{r}  \right |^{2}$) based on the PS technique which is given by
	\begin{align}\label{eq8}
		E^{PS}=\frac{\chi[\varepsilon\left ( 1-\rho  \right )\mathcal{T}]}{2}\eta_{s}\left | g_{sr}\right |^{2}w_{sr}^{-2}.
	\end{align}
	Hence, the resulting information signal (sent by $S$) at $R$ after processing in this phase can be expressed as
	\begin{align}\label{eq9}
		y_{r}^{IP}&=\sqrt{1-\varepsilon }\big(\sqrt{\eta_s w_{sr}^{-2}}g_{sr}x_{s}+n_{r}\big)+ n_{rb},
	\end{align}
	where $n_{rb}\sim \mathbb{CN}\left(0,\sigma_{rb}^{2}\right)$ is RF-to-baseband conversion noise. 
	
	In the third phase (designated as IP Phase II), the ATx $R$ first amplifies the signal $y_{r}^{IP}$ by a gain factor\footnote{As such, using (\ref{eq9}), the AF-based gain at $R$ can be calculated as $\mathcal{G}= (1-\varepsilon)({\eta_s}|g_{sr}|^2w_{sr}^{-2}+\sigma_r^2)+\sigma_{rb}^2$. However, the exact value of $\mathcal{G}$ makes the forthcoming system performance analysis cumbersome. Therefore, we employ an approximated gain in this work, i.e., $\mathcal{G}\approx(1-\varepsilon){\eta_s}|g_{sr}|^2w_{sr}^{-2}$, where the noise power terms are omitted. It may be justifiable owing to the fact that the magnitude of noise power is negligible as compared to the signal power.} $\mathcal{G}$, and then, superposes the resulting signal with its own information $x_r$ by multiplexing the two signals in power domain. In particular, the net available power $\mathcal{P}_{h}$ at $R$ (after energy harvesting in previous phases) is split into two parts, i.e., $\mu\mathcal{P}_{h}$ and $(1-\mu)\mathcal{P}_{h}$ to generate a network-coded signal as 
	\begin{align}\label{eq10}
		x_{s,r}&=\sqrt{\frac{\mu \mathcal{P}_{h}}{\mathcal{G}}}y_{r}^{IP}+\sqrt{\left( 1-\mu \right)\mathcal{P}_{h}}x_{r},
	\end{align}
	where $\mu\in(0,1)$ is referred to as a spectrum sharing factor since it facilitates the spectrum sharing between satellite and A2A networks.
	Herein, we consider a practical non-linear energy harvester at $R$, for which, the generated power $\mathcal{P}_{h}$ can be calculated using (\ref{eq7}) and (\ref{eq8}) by adopting the piecewise linear EH model \cite{51}, \cite{52} as

	\begin{equation}\label{eq15}
		\mathcal{P}_{h} =\left\{\!\begin{array}{ll}
			\chi_{\rho,\varepsilon}\eta_s\left | g_{sr}\right |^{2} w_{sr}^{-2}, &{\text{for }}\,{\eta_s \left | g_{sr}\right |^{2}}w_{sr}^{-2} \leq {\mathcal{P}_\text{th}},\\
			\chi_{\rho,\varepsilon}\mathcal{P}_\text{th}, &{\text{for }}\,{\eta_s \left | g_{sr}\right |^{2}}w_{sr}^{-2} >{\mathcal{P}_\text{th}},
		\end{array}\right.
	\end{equation}
	where $\chi_{\rho,\varepsilon}\triangleq\chi( {2\rho }/{(1-\rho) } +\varepsilon )$ and $\mathcal{P}_\text{th}$ is the saturation threshold power. In all the expressions derived in this paper, the corresponding expressions for linear EH can be obtained simply by substituting $\mathcal{P}_\text{th}\to \infty$. Hereby, we assume that all the harvested energy at $R$ is available for the transmission purpose with negligible energy consumption by signal processing circuitry. Finally, in this phase, the $R$ broadcasts the superposed signal $x_{s,r}$ which is received by $D_i$ and $T$. The received signals at $D_i$ and $T$ can be compactly expressed, respectively, as
	\begin{align}\label{16}  
		y_{d_i}\!&=\!w_{rd_i}^{-\nu_{rd_i}/2}g_{rd_i}\left(\sqrt{\frac{\mu \mathcal{P}_{h}}{\mathcal{G}}}y_{r}^{IP}+\sqrt{\left( 1-\mu \right)\mathcal{P}_{h}}x_{r}\right)+n_{d_i},
	\end{align} 
	and 
	\begin{align}\label{eq12}  
		y_{t}\!&=\!w_{rt}^{-\nu_{rt}/2}g_{rt}\left(\sqrt{\frac{\mu \mathcal{P}_{h}}{\mathcal{G}}}y_{r}^{IP}+\sqrt{\left( 1-\mu \right)\mathcal{P}_{h}}x_{r}\right)+n_{t},
	\end{align} 
	where $n_{d_i}\sim \mathbb{CN}\left( 0,\sigma_{d_i}^{2} \right)$ and $n_{t}\sim \mathbb{CN}\left( 0,\sigma_{t}^{2} \right)$ are the AWGNs for respective links. On substituting the $\mathcal{P}_{h}$ as given by (\ref{eq15}) in (\ref{16}), the signal-to-noise ratio (SNR) at $D_i$ via $R$ can be calculated as
	\begin{equation}\label{17}
		\Lambda_{D_i} =\left\{\!\begin{array}{ll}
			\Lambda_{D_i,1}, &{\text{for }}\,{\eta_s \left | g_{sr}\right |^{2}}w_{sr}^{-2} \leq {\mathcal{P}_\text{th}},\\
			\Lambda_{D_i,2}, &{\text{for }}\,{\eta_s \left | g_{sr}\right |^{2}}w_{sr}^{-2} >{\mathcal{P}_\text{th}},
		\end{array}\right.
	\end{equation}
	where $\Lambda_{D_i,1}$ and $\Lambda_{D_i,2}$ are given, respectively, as
		\begin{align}\label{eq23} 
			\Lambda_{D_i,1}&=\frac{\mu\chi_{\rho,\varepsilon}\eta_s\left| g_{sr} \right|^{2}w_{sr}^{-2}\left| g_{rd_i} \right|^{2}w^{-\nu_{rd_i}}_{rd_i}}{\mu_\varepsilon\chi_{\rho,\varepsilon} \left| g_{rd_i} \right|^{2}w^{-\nu_{rd_i}}_{rd_i}+\left( 1-\mu \right)\chi_{\rho,\varepsilon} \eta_s\left| g_{sr} \right|^{2}w_{sr}^{-2}\left| g_{rd_i} \right|^{2}w^{-\nu_{rd_i}}_{rd_i}+\sigma^{2}_{d_{i}}}
		\end{align}
and
		\begin{align}\label{eq24}
			\Lambda_{D_i,2}&=\frac{\mu\chi_{\rho,\varepsilon}\mathcal{P}_\text{th}\eta_s\left| g_{sr} \right|^{2}w_{sr}^{-2}\left| g_{rd_i} \right|^{2}w^{-\nu_{rd_i}}_{rd_i}}{\mu_\varepsilon\chi_{\rho,\varepsilon}\mathcal{P}_\text{th}\left| g_{rd_i} \right|^{2}w^{-\nu_{rd_i}}_{rd_i}+\left( 1-\mu \right)\chi_{\rho,\varepsilon}\eta_s\mathcal{P}_\text{th}  \left| g_{sr} \right|^{2}w_{sr}^{-2}\left| g_{rd_i} \right|^{2}w^{-\nu_{rd_i}}_{rd_i}+\eta_s\left| g_{sr} \right|^{2}w_{sr}^{-2}\sigma^{2}_{d_{i}}}
		\end{align}
containing $\mu_\varepsilon \triangleq\mu\left (   \sigma _{r}^{2}+{\sigma _{rb}^{2}}/{( 1-\varepsilon) } \right )$.	

	Further, from (\ref{eq12}), we observe that the received signal at $T$ contains a large interference term due to the satellite signal component $y_{r}^{IP}$.	For reliable A2A communications, the ARx node $T$ must be equipped with the ability to eliminate this interference by making use of the copy of satellite signal overheard by the $T$ along with the $R$ in the first two phases of the transmission block (or by employing a multiuser decoder). However, in practice, the cancellation of this interference is difficult due to the requirement of very high SNR and/or very good channel conditions, e.g., clear sky, etc. over the link between $S$ and $T$. Thus, based on the practical imperfect interference cancellation (im-IC) in (\ref{eq12}) and thereby substituting the $\mathcal{P}_{h}$ using (\ref{eq15}), the SNR at $T$ can be obtained as  
	\begin{equation}\label{21}
		\Lambda^\text{im-IC}_{T} =\left\{\!\begin{array}{ll}
			\Lambda^\text{im-IC}_{T,1}, &{\text{for }}\,{\eta_s \left | g_{sr}\right |^{2}}w_{sr}^{-2} \leq {\mathcal{P}_\text{th}},\\
			\Lambda^\text{im-IC}_{T,2}, &{\text{for }}\,{\eta_s \left | g_{sr}\right |^{2}}w_{sr}^{-2} >{\mathcal{P}_\text{th}},
		\end{array}\right.
	\end{equation}
	where $\Lambda^\text{im-IC}_{T,1}$ and $\Lambda^\text{im-IC}_{T,2}$ are shown, respectively, as 	
		\begin{align}\label{22} 
			\Lambda^\text{im-IC}_{T,1}=\frac{(1-\mu)\chi_{\rho,\varepsilon}\eta_s\left| g_{sr} \right|^{2}w_{sr}^{-2}\left| g_{rt} \right|^{2}w^{-\nu_{rt}}_{rt}}{\mu_\varepsilon \chi_{\rho,\varepsilon}\left| g_{rt} \right|^{2}w^{-\nu_{rt}}_{rt}+ \mu \chi_{\rho,\varepsilon} \eta_s\left| g_{sr} \right|^{2}w_{sr}^{-2} \left| g_{rt} \right|^{2}w^{-\nu_{rt}}_{rt}+\sigma^{2}_{t}} 
		\end{align}
	and
		\begin{align}\label{23} 
			\Lambda^\text{im-IC}_{T,2}=\frac{(1-\mu)\chi_{\rho,\varepsilon}\mathcal{P}_\text{th}\eta_s\left| g_{sr} \right|^{2}w_{sr}^{-2}\left| g_{rt} \right|^{2}w^{-\nu_{rt}}_{rt}}{\mu_\varepsilon \chi_{\rho,\varepsilon}\mathcal{P}_\text{th}\left| g_{rt} \right|^{2}w^{-\nu_{rt}}_{rt}+ \mu \chi_{\rho,\varepsilon}\mathcal{P}_\text{th} \eta_s\left| g_{sr} \right|^{2}w_{sr}^{-2} \left| g_{rt} \right|^{2}w^{-\nu_{rt}}_{rt}+\eta_s\left| g_{sr} \right|^{2}w_{sr}^{-2}\sigma^{2}_{t}}. 
		\end{align}
	Although the case of perfect interference cancellation (p-IC) appears to be idealistic, it is important to consider this scenario for the benchmarking purpose of the system performance. The resulting SNR under this scenario can be expressed as   
	\begin{equation}\label{24}
		\Lambda^\text{p-IC}_{T} =\left\{\!\begin{array}{ll}
			\Lambda^\text{p-IC}_{T,1}, &{\text{for }}\,{\eta_s \left | g_{sr}\right |^{2}}w_{sr}^{-2} \leq {\mathcal{P}_\text{th}},\\
			\Lambda^\text{p-IC}_{T,2}, &{\text{for }}\,{\eta_s \left | g_{sr}\right |^{2}}w_{sr}^{-2} >{\mathcal{P}_\text{th}},
		\end{array}\right.
	\end{equation}
	where 
		\begin{align}\label{n25} 
			\Lambda^\text{p-IC}_{T,1}=\frac{(1-\mu)\chi_{\rho,\varepsilon}\eta_s\left| g_{sr} \right|^{2}w_{sr}^{-2}\left| g_{rt} \right|^{2}w^{-\nu_{rt}}_{rt}}{\mu_\varepsilon \chi_{\rho,\varepsilon}\left| g_{rt} \right|^{2}w^{-\nu_{rt}}_{rt}+ \sigma^{2}_{t}} 
		\end{align}
		and
		\begin{align}\label{n26} 
			\Lambda^\text{p-IC}_{T,2}=\frac{(1-\mu)\chi_{\rho,\varepsilon}\mathcal{P}_\text{th}\eta_s\left| g_{sr} \right|^{2}w_{sr}^{-2}\left| g_{rt} \right|^{2}w^{-\nu_{rt}}_{rt}}{\mu_\varepsilon \chi_{\rho,\varepsilon}\mathcal{P}_\text{th}\left| g_{rt} \right|^{2}w^{-\nu_{rt}}_{rt}+\eta_s\left| g_{sr} \right|^{2}w_{sr}^{-2}\sigma^{2}_{t}}. 
		\end{align}
		
		
		\section{Outage Performance of S2G Network}\label{eop}
		In this section, we evaluate the OP of the S2G network of considered OSAGIN under SR and Nakagami-\emph{m} fading for the satellite and ground links, respectively, by taking into account the corresponding distance distributions.
		
		For a threshold $\gamma_{\textmd{S}}$, where $\gamma_{\textmd{S}}=2^{\frac{2r_{s}}{1-\rho}}-1$ and $r_{s}$ is a target rate, the OP of the S2G network when $S$ communicates with a typical GU $D_i$ via ATx $R$ is given by
		\begin{align}\label{eq21}
			\mathcal{P}^{\textmd{S2G}}_{\textmd{out}}(\gamma_\textmd{S}) 
			&=\textmd{Pr}\left[\Lambda_{D_i}<\gamma_\textmd{S}\right],
		\end{align}
		which can be re-expressed (based on (\ref{17})) as
		\begin{align}\label{27}
			\mathcal{P}^{\textmd{S2G}}_{\textmd{out}}(\gamma_\textmd{S}) 
			&= 1-\underbrace{\textmd{Pr}\left[ \Lambda_{D_i,1}  > \gamma_\textmd{S},\eta_s\left| g_{sr} \right|^{2}w_{sr}^{-2}\leq \mathcal{P}_\text{th} \right] }_{	\mathcal{P}_1(\gamma_\textmd{S})}-\underbrace{\textmd{Pr}\left[ \Lambda_{D_i,2}  > \gamma_\textmd{S},\eta_s\left| g_{sr} \right|^{2}w_{sr}^{-2}> \mathcal{P}_\text{th} \right] }_{	\mathcal{P}_2(\gamma_\textmd{S})},
			\end{align}
		In (27), we need to evaluate the joint probability terms $\mathcal{P}_1(\gamma_\textmd{S})$ and $\mathcal{P}_2(\gamma_\textmd{S})$ which is quite challenging due to the involvement of multiple random variables corresponding to random distances and multipath fading channels related to underlying links. 
		We proceed with the derivation of $\mathcal{P}_1(\gamma_\textmd{S})$ and $\mathcal{P}_2(\gamma_\textmd{S})$ in the following theorems.
		\newtheorem{theorem}{Theorem}
		\begin{theorem}	\label{theo1} 
			The probability $\mathcal{P}_1(\gamma_\textmd{S})$ in (\ref{27}) under i.i.d. links is given by
			\begin{equation}\label{32}
				\mathcal{P}_1(\gamma_\textmd{S})=\left\{\!\begin{array}{ll}
					\Psi_1(\gamma_{\textmd{S}}), &{\text{for }}\gamma_{\textmd{S}}< \mu^\prime \mbox{ and } {\mathcal{P}_\text{th}}\geq\left(\frac{\mathcal{B}}{\mathcal{A}}\right){\eta_s},\\
					0, &{\text{for }}\gamma_{\textmd{S}}< \mu^\prime \mbox{ and } {\mathcal{P}_\text{th}}<\left(\frac{\mathcal{B}}{\mathcal{A}}\right){\eta_s},\\
					0, &{\text{for }}\gamma_{\textmd{S}}\geq \mu^\prime,
				\end{array}\right.
			\end{equation}
			where $\Psi_{\textmd{1}}(\gamma_\textmd{S})$ is shown as 
				\begin{align}\label{26}
					\Psi_{\textmd{1}}(\gamma_\textmd{S})
					&=\frac{\alpha _{sr}\mathcal{A}}{l^2~w_{er}w_{\min}\nu_{rd_i}} \sum _{k =0}^{m_{sr}-1}\zeta\left ({k }\right)\sum_{k_1=0}^{k}\binom{k}{k_1}\sum_{k_2=0}^{\infty }\frac{(-1)^{k_2}\mathcal{B}^{-k_1-k_2-2}}{\bar{\beta}_{sr}^{k+2}k_2!}\nonumber\\
					&\times\Delta_\Gamma\left( k+k_2+2,\frac{\bar{\beta}_{sr}\mathcal{B}w_{\min}^{2}}{\mathcal{A}},\frac{\bar{\beta}_{sr}\mathcal{B}w_{\max}^{2}}{\mathcal{A}} \right)\sum_{n=0}^{m_{rd_i}-1}{\mathcal{P}_{\mathcal{A}, \mathcal{B}}}^{\varsigma_1} \frac{\left( m_{rd_i}\sigma^2_{d_i}\gamma_\textmd{S}\right)^n}{n!}\nonumber\\
					&\times\left[(h^2_0+l^2)^{\frac{\nu_{rd_i}n+2}{2}} G^{2,1}_{2,3}
					\bigg[\frac{m_{rd_i}\sigma^2_{d}\gamma_\textmd{S}(h^2_0+l^2)^{\nu_{rd_i}/2}}{{\mathcal{P}_{\mathcal{A}, \mathcal{B}}}} \bigg\vert{{1-n-\frac{2}{\nu_{rd_i}},\varsigma_1+1}\atop{{\varsigma_1,0,-n-\frac{2}{\nu_{rd_i}}}}}\bigg] \right. \nonumber\\
					&\left. -h_0^{{\nu_{rd_i}n+2}} G^{2,1}_{2,3}
					\bigg[\frac{m_{rd_i}\sigma^2_{d}\gamma_\textmd{S}h_0^{{\nu_{rd_i}}}}{{\mathcal{P}_{\mathcal{A}, \mathcal{B}}}} \bigg\vert{{1-n-\frac{2}{\nu_{rd_i}},\varsigma_1+1}\atop{{\varsigma_1,0,-n-\frac{2}{\nu_{rd_i}}}}}\bigg]\right]
				\end{align}  
		\end{theorem}	
		\begin{IEEEproof}
			Please refer to the Appendix B.
		\end{IEEEproof}
			\begin{theorem}	\label{theo2}
				The probability $\mathcal{P}_2(\gamma_\textmd{S})$ in (\ref{27}) is given by
				\begin{equation}\label{36}
					\mathcal{P}_2(\gamma_\textmd{S})=\left\{\!\begin{array}{ll}
						\Psi_2(\gamma_{\textmd{S}}), &{\text{for }}\gamma_{\textmd{S}}< \mu^\prime \mbox{ and } {\mathcal{P}_\text{th}}\geq\left(\frac{\mathcal{B}}{\mathcal{A}}\right){\eta_s},\\
						\Psi_3(\gamma_{\textmd{S}}), &{\text{for }}\gamma_{\textmd{S}}< \mu^\prime \mbox{ and } {\mathcal{P}_\text{th}}<\left(\frac{\mathcal{B}}{\mathcal{A}}\right){\eta_s},\\
						0, &{\text{for }}\gamma_{\textmd{S}}\geq \mu^\prime,
					\end{array}\right.
				\end{equation}
				where $\Psi_{\textmd{2}}(\gamma_\textmd{S})$ is
					\begin{align}\label{e40}
						\Psi_{2}(\gamma_\textmd{S})	&=\frac{2\alpha _{sr}}{ l^2w_{er}w_{\min}\nu_{rd_i}} \left(\frac{m_{rd_i}\sigma^2_{d_i}\gamma_\textmd{S}\eta_{s}}{\mathcal{P}_\text{th} }\right)^{-\frac{2}{\nu_{rd_i}}}\sum _{k =0}^{m_{sr}-1}  \zeta\left ({k }\right) \nonumber\\
						&\times\sum_{n=0}^{m_{rd_i}-1}\frac{1}{n!}\sum_{k_1=0}^{k+n}\binom{k+n}{k_1}\sum_{k_2=0}^{\infty }	\frac{(-1)^{k_2}}{k_2!}{\bar{\beta}_{sr}}^{\frac{n-k_1+k_2}{2}-1} \frac{\mathcal{P}_{\mathcal{A,B}}^{\frac{-n+k_1-k_2}{2}+1}\mathcal{B}^{k+n-k_1+k_2}}{\mathcal{A}^{k+1+\frac{n-k_1-k_2+2}{2}}-\frac{2}{\nu_{rd_i}}}\nonumber\\
						&\times
						\Delta_\Gamma\left( n+k_2+\frac{2}{\nu_{rd_i}},\frac{m_{rd}\sigma^2_{d_i}\gamma_\textmd{S}\eta_{s} h_0^{\nu_{rd_i}}}{\mathcal{A}\mathcal{P}_\text{th} },\frac{m_{rd}\sigma^2_{d_i}\gamma_\textmd{S}\eta_{s} (h^2_0+l^2)^{\frac{\nu_{rd_i}}{2}}}{\mathcal{A}\mathcal{P}_\text{th} } \right)\mathbb{I}_1
					\end{align}
with $\mathbb{I}_1$ as
					\begin{align}\label{40aa}
						\mathbb{I}_1&=\int_{w_{\min}}^{w_{\max}}  w^{2k+n-k_1+k_2+1}\exp\left(-\frac{\bar{\beta}_{sr}w^2}{\mathcal{A}} \left(\mathcal{B}+\frac{\mathcal{P}_{\mathcal{A,B}}}{2}\right) \right)\mathbb{W}_{\frac{-n+k_1-k_2}{2},\frac{+n-k_1+k_2-1}{2}}\left(\frac{\bar{\beta}_{sr}w^2 \mathcal{P}_{\mathcal{A,B}}}{\mathcal{A}} \right)dw.
					\end{align}
				Further, $\Psi_{\textmd{3}}(\gamma_\textmd{S})$ is  
					\begin{align}\label{e41}
						\Psi_{\textmd{3}}(\gamma_\textmd{S})
						&=\frac{2\alpha _{sr}}{ l^2w_{er}w_{\min}\nu_{rd_i}} \sum _{k =0}^{m_{sr}-1}  \zeta\left ({k }\right)\sum_{n=0}^{m_{rd_i}-1} \sum_{k_{1}=0}^{k+n}\binom{k+n}{k_{1}}\nonumber\\
						&\times\sum _{k_2 =0}^{\infty} \frac{(-1)^{k_2}}{n!~k_2!}\frac{\bar{\beta}_{sr}^{\frac{n-k_1-1}{2}}{ \mathcal{B}} ^{\frac{2k+n-k_1+1}{2}}}{\mathcal{A}^{k+n+1+k_2}} \left(\frac{m_{rd_i}\sigma^2_{d_i}\eta_s\gamma_\textmd{S}}{\mathcal{P}_\textmd{th}}  \right)^{\frac{n+k_1+1}{2}+k_2}	\mathbb{I}_2 
					\end{align}
with $\mathbb{I}_2$ as 				
					\begin{align}\label{e41a}
						&\mathbb{I}_2=\int_{w_{\min}}^{w_{\max}} \left[ {\left( h^2_0+l^2 \right)}^{\frac{\nu_{rd_i}\varsigma_2}{2}+1} G^{2,1}_{1,3}
						\bigg[\frac{m_{rd_i}\sigma^2_{d_i}\eta_s\gamma_\textmd{S}\left( h^2_0+l^2 \right)^{\nu_{rd_i}/2}\mathcal{B}\bar{\beta}_{sr}w^2}{\mathcal{A}^{2}\mathcal{P}_\textmd{th}} \bigg\vert{{1-\varsigma_2-\frac{2}{\nu_{rd_i}}}\atop{{\varsigma_3,-\varsigma_3}},-\varsigma_2-\frac{2}{\nu_{rd_i}}}\bigg]\right. \nonumber\\
						&\left.- h_0^{\nu_{rd_i}\varsigma_2+2} G^{2,1}_{1,3}
						\bigg[\frac{m_{rd_i}\sigma^2_{d_i}\eta_s\gamma_\textmd{S}h_0^{\nu_{rd_i}}\mathcal{B}\bar{\beta}_{sr}w^2}{\mathcal{A}^{2}\mathcal{P}_\textmd{th}} \bigg\vert{{1-\varsigma_2-\frac{2}{\nu_{rd_i}}}\atop{{\varsigma_3,-\varsigma_3}},-\varsigma_2-\frac{2}{\nu_{rd_i}}}\right]  w^{2k+n-k_1+2}\exp\left( -\frac{\mathcal{B}\bar{\beta}_{sr}w^2}{\mathcal{A}} \right)dw.
					\end{align}
			\end{theorem}	
			\begin{IEEEproof}
				Please refer to the Appendix C.
			\end{IEEEproof}
			Here, $\mathbb{I}_1$ and $\mathbb{I}_2$ can be evaluated using the Chebyshev-Gauss Quadrature (CGQ) method. CGQ is a numerical integration method that approximates a definite integral of a function $f(w)$ of variable $w\in[w_{\min}, w_{\max}]$ as 
			\begin{align}\label{cgq}
				\int_{w_{\min}}^{w_{\max}}f(w)dw&\stackrel{(a)}{=}\int_{-1}^{1}b_1f(b_1\tau+b_2)d\tau \stackrel{(b)}{\approx} \frac{\pi}{n}\sum_{i=1}^{n}\sqrt{1-\varpi _{i}^{2}}f(b_1\varpi _{i}+b_2),
			\end{align}
			where $b_1=\frac{w_{\max}-w_{\min}}{2}$, $b_2=\frac{w_{\max}+w_{\min}}{2}$, $\varpi _{i}=\cos\left( \frac{2i-1}{2n} \pi\right)$, and $n$ is a complexity-accuracy trade-off parameter. In (\ref{cgq}), the step in ($a$) involves a change of variable $w=b_1\tau+b_2$ followed by the step ($b$) that applies \cite[eq 25.4.38]{formula}. 
			\section{Outage Performance of A2A Network}\label{eops}
			In this section, we evaluate the OP of the A2A network of considered OSAGIN when the A2A link undergoes Rician fading. Hereby, the ARx is randomly located inside the beam-inspired 3D conical coverage region below ATx. Both the im-IC and p-IC scenarios at ARx will be considered for the evaluation of OP of the A2A network. 
			
			\subsection{With Imperfect Interference Cancellation}
			 In this section, we evaluate the OP when the ARx can successfully decode the  signal in the first phase. Consequently,  $T$  can eliminate the interference from  signal in the second IP phase. For a target threshold $\gamma_{\textmd{A}}$,  where $\gamma_{\textmd{A}}=2^{\frac{2r_{a}}{1-\rho}}-1$ and $r_{a}$ is a target rate, the OP of the A2A network, where ATx $R$ communicates with ARx $T$ under im-IC is given by
			\begin{align}\label{eq31}
				\mathcal{P}^{\textmd{A2A}}_{\textmd{out}}(\gamma_\text{A}) 
				&=\textmd{Pr}\left[\Lambda^\text{im-IC}_{T}<\gamma_\text{A}\right],
			\end{align}
			which can be re-expressed according to (\ref{21}) as
			\begin{align}\label{eq32}
				\mathcal{P}^{\textmd{A2A}}_{\textmd{out}}(\gamma_\text{A}) 
				&= 1-\underbrace{\textmd{Pr}\left[\Lambda^\text{im-IC}_{T,1}  > \gamma_\text{A},\eta_s\left| g_{sr} \right|^{2}w_{sr}^{-2}\leq \mathcal{P}_\text{th} \right] }_{	\mathcal{P}_3(\gamma_\text{A})}-\underbrace{\textmd{Pr}\left[ \Lambda^\text{im-IC}_{T,2}  > \gamma_\text{A},\eta_s\left| g_{sr} \right|^{2}w_{sr}^{-2}> \mathcal{P}_\text{th} \right] }_{	\mathcal{P}_4(\gamma_\text{A})},
			\end{align}
			where $\Lambda^\text{im-IC}_{T,1}$ and $\Lambda^\text{im-IC}_{T,2}$ are given as (\ref{22}) and (\ref{23}), respectively. The joint probability terms $	\mathcal{P}_3(\gamma_\text{A})$ and 	$\mathcal{P}_4(\gamma_\text{A})$ are derived in the following theorems. 
			\begin{theorem}	\label{theo4}
				The probability $\mathcal{P}_{3}(\gamma_\text{A})$ in (\ref{eq32}) is given by
				\begin{equation}\label{eq35}
					\mathcal{P}_3(\gamma_\textmd{A})=\left\{\!\begin{array}{ll}
						\widetilde{\Psi}_1(\gamma_\textmd{A}), &{\text{for }}\gamma_{\textmd{A}}< 1/\mu^\prime \mbox{ and } {\mathcal{P}_\text{th}}\geq\left(\frac{\mathcal{D}}{\mathcal{C}}\right){\eta_s},\\
						0, &{\text{for }}\gamma_{\textmd{A}}< 1/\mu^\prime \mbox{ and } {\mathcal{P}_\text{th}}<\left(\frac{\mathcal{D}}{\mathcal{C}}\right){\eta_s},\\
						0, &{\text{for }}\gamma_{\textmd{A}}\geq 1/\mu^\prime,
					\end{array}\right.
				\end{equation}
				where 
				$\widetilde{\Psi}_1(\gamma_\textmd{A})$ is shown as   
					\begin{align}\label{eq36}
						\widetilde{\Psi}_1(\gamma_\textmd{A})&= \frac{3\mathcal{C}\alpha _{sr}}{w_{er}w_{\min}\nu_{rt}\left( {h_{2}^{3}-h_{1}^{3}} \right)} \sum _{k =0}^{m_{sr}-1} \frac{\zeta\left ({k }\right)\exp(-K_{rt})}{\bar{\beta}^{k+2}_{sr}} \sum_{m=0}^{\infty }\sum_{n=0}^{m}\frac{K_{rt}^{m }[\Theta(\gamma_\textmd{A})]^{n}  }{m!n!}
						\nonumber\\
						&\times\sum_{k_1=0}^{k}\binom{k}{k_1} \sum_{k_2=0}^{\infty }\frac{(-1)^{k_2}}{k_2!} 	 \frac{\mathcal{P}_{\mathcal{C},\mathcal{D}}^{\varsigma_1}}{\mathcal{D}^{k_1+k_2+2}}\Delta_\Gamma\left( k+k_2+2,\frac{\bar{\beta}_{sr}\mathcal{D}w_{\min}^{2}}{\mathcal{C}},\frac{\bar{\beta}_{sr}\mathcal{D}w_{\max}^{2}}{\mathcal{C}} \right)\nonumber\\
						&\times \left[h_{2}^{{\nu_{rt}n+3}}G^{2,1}_{2,3}
						\bigg[\frac{\Theta(\gamma_\textmd{A})h_{2}^{\nu_{rt}}}{\mathcal{P}_{\mathcal{C}, \mathcal{D}}}\bigg\vert{{1-n-\frac{3}{\nu_{rt}},\varsigma_1+1}\atop{{\varsigma_1,0,-n-\frac{2}{\nu_{rt}}}}}\bigg]\right.\nonumber\\
						&-h_{1}^{{\nu_{rt}n+3}} G^{2,1}_{2,3}
						\bigg[\frac{\Theta(\gamma_\textmd{A})h_{1}^{\nu_{rt}}}{\mathcal{P}_{\mathcal{C}, \mathcal{D}}}\bigg\vert{{1-n-\frac{3}{\nu_{rt}},\varsigma_1+1}\atop{{\varsigma_1,0,-n-\frac{2}{\nu_{rt}}}}}\bigg]\nonumber\\
						&-h_{1}\left( \frac{h_{2}}{\cos \upphi } \right)^{{\nu_{rt}n+2}} G^{2,1}_{2,3}
						\bigg[\frac{\Theta(\gamma_\textmd{A})h_{2}^{\nu_{rt}}}{\mathcal{P}_{\mathcal{C}, \mathcal{D}}{\cos^{\nu_{rt}} \upphi }}\bigg\vert{{1-n-\frac{2}{\nu_{rt}},\varsigma_1+1}\atop{{\varsigma_1,0,-n-\frac{2}{\nu_{rt}}}}}\bigg]\nonumber\\
						& +h_{1}^{{\nu_{rt}n+3}} G^{2,1}_{2,3}
						\bigg[\frac{\Theta(\gamma_\textmd{A})h_{1}^{\nu_{rt}}}{\mathcal{P}_{\mathcal{C}, \mathcal{D}}}\bigg\vert{{1-n-\frac{2}{\nu_{rt}},\varsigma_1+1}\atop{{\varsigma_1,0,-n-\frac{2}{\nu_{rt}}}}}\bigg]\nonumber\\
						&-{\cos \upphi } \left( \frac{h_{2}}{\cos \upphi } \right)^{{\nu_{rt}n+3}}G^{2,1}_{2,3}
						\bigg[\frac{\Theta(\gamma_\textmd{A})h_{2}^{\nu_{rt}}}{\mathcal{P}_{\mathcal{C}, \mathcal{D}}{\cos^{\nu_{rt}} \upphi }}\bigg\vert{{1-n-\frac{3}{\nu_{rt}},\varsigma_1+1}\atop{{\varsigma_1,0,-n-\frac{3}{\nu_{rt}}}}}\bigg]\nonumber\\
						&+\cos \upphi  \left( \frac{h_{1}}{\cos \upphi } \right)^{{\nu_{rt}n+3}} G^{2,1}_{2,3}
						\bigg[\frac{\Theta(\gamma_\textmd{A})h_{1}^{\nu_{rt}}}{\mathcal{P}_{\mathcal{C}, \mathcal{D}}{\cos \upphi }}\bigg\vert{{1-n-\frac{3}{\nu_{rt}},\varsigma_1+1}\atop{{\varsigma_1,0,-n-\frac{3}{\nu_{rt}}}}}\bigg]\nonumber\\
						&+h_2\left( \frac{h_{2}}{\cos \upphi } \right)^{{\nu_{rt}n+2}}G^{2,1}_{2,3}
						\bigg[\frac{\Theta(\gamma_\textmd{A})h_{2}^{\nu_{rt}}}{\mathcal{P}_{\mathcal{C}, \mathcal{D}}{\cos \upphi }}\bigg\vert{{1-n-\frac{2}{\nu_{rt}},\varsigma_1+1}\atop{{\varsigma_1,0,-n-\frac{2}{\nu_{rt}}}}}\bigg]\nonumber\\ 
						&\left. -h_{2}^{{\nu_{rt}n+3}} G^{2,1}_{2,3}
						\bigg[\frac{\Theta(\gamma_\textmd{A})h_{2}^{\nu_{rt}}}{\mathcal{P}_{\mathcal{C}, \mathcal{D}}}\bigg\vert{{1-n-\frac{2}{\nu_{rt}},\varsigma_1+1}\atop{{\varsigma_1,0,-n-\frac{2}{\nu_{rt}}}}}\bigg]\right]
					\end{align}
			with $ \Theta(\gamma_\textmd{A})= {(1+K_{rt})\sigma^{2}_{t}\gamma_\textmd{A}} $. 
			\end{theorem}	
			\begin{IEEEproof} 	
				Please refer to the Appendix D.
			\end{IEEEproof}
			
			\begin{theorem}	\label{theo3}
				The probability $\mathcal{P}_{4}(\gamma_\text{A})$ in (\ref{eq32}) is given by,
				\begin{equation}\label{eq35b}
					\mathcal{P}_4(\gamma_\textmd{A})=\left\{\!\begin{array}{ll}
						\widetilde{\Psi}_2(\gamma_\textmd{A}), &{\text{for }}\gamma_{\textmd{A}}< 1/\mu^\prime \mbox{ and } {\mathcal{P}_\text{th}}\geq\left(\frac{\mathcal{D}}{\mathcal{C}}\right){\eta_s},\\
						\widetilde{\Psi}_3(\gamma_\textmd{A}), &{\text{for }}\gamma_{\textmd{A}}< 1/\mu^\prime \mbox{ and } {\mathcal{P}_\text{th}}<\left(\frac{\mathcal{D}}{\mathcal{C}}\right){\eta_s},\\
						0, &{\text{for }}\gamma_{\textmd{A}}\geq 1/\mu^\prime,
					\end{array}\right.
				\end{equation}
				where $\widetilde{\Psi}_2(\gamma_\textmd{A})$ is 
					\begin{align}\label{eq30a}
						\widetilde{\Psi}_{2}(\gamma_\textmd{A})&=\frac{6\alpha _{sr}}{\nu_{rt}\left( {h_{2}^{3}-h_{1}^{3}} \right)} \sum _{k =0}^{m_{sr}-1}  \zeta\left ({k }\right)\sum_{m=0}^{\infty }\sum_{n=0}^{m}\frac{K_{rt}^{m} \exp(-K_{rt})}{m! n!}\sum_{k1=0}^{n+k}\binom{n+k}{k_1}\nonumber\\
						&\times\sum_{k_2=0}^{\infty }\frac{(-1)^{k_2}}{k_2!}\left( \frac{\eta_s{\Theta(\gamma_{A})}}{\mathcal{P}_\text{th}} \right)^{-2/\nu_{rt}}{\bar{\beta}_{sr}}^{\frac{n-k_1+k_2}{2}+1}\mathcal{P}_{\mathcal{C,D}}^{\frac{-n+k_1-k_2}{2}}\frac{\mathcal{D}^{k+p_1-k_1+k_2}}{ w_{er}w_{\min}\mathcal{C}^{k+2\frac{p_1-k_1+k_2}{2}}-\frac{2}{\nu_{rt}}}\nonumber\\
						&\times\left [\left( \frac{\eta_s{\Theta(\gamma_{A})}}{\mathcal{P}_\text{th}} \right)^{-1/\nu_{rt}}\Delta_\Gamma\left( p_1+k_2+\frac{3}{\nu_{rt}},\frac{\eta_s{\Theta(\gamma_{A})h_1}}{\mathcal{C}\mathcal{P}_\text{th}},\frac{\eta_s{\Theta(\gamma_{A})h_2}}{\mathcal{C}\mathcal{P}_\text{th}} \right)\right.\nonumber\\
						&- h_1\Delta_\Gamma\left( p_1+k_2+\frac{2}{\nu_{rt}},\frac{\eta_s{\Theta(\gamma_{A})h_1}}{\mathcal{C}\mathcal{P}_\text{th}},\frac{\eta_s{\Theta(\gamma_{A})h_1}}{\mathcal{C}\mathcal{P}_\text{th}\cos \upphi} \right)\nonumber\\
						&\left.+h_2\Delta_\Gamma\left( p_1+k_2+\frac{2}{\nu_{rt}},\frac{\eta_s{\Theta(\gamma_{A})h_2}}{\mathcal{C}\mathcal{P}_\text{th}\cos \upphi},\frac{\eta_s{\Theta(\gamma_{A})h_2}}{\mathcal{C}\mathcal{P}_\text{th}} \right)\right. \nonumber\\
						&- \left.{\left( \frac{\eta_s{\Theta(\gamma_{A})}}{\mathcal{P}_\text{th}} \right)^{-1/\nu_{rt}}\cos \upphi \Delta_\Gamma\left( p_1+k_2+\frac{3}{\nu_{rt}},\frac{\eta_s{\Theta(\gamma_{A})h_2}}{\mathcal{C}\mathcal{P}_\text{th}\cos \upphi},\frac{\eta_s{\Theta(\gamma_{A})h_1}}{\mathcal{C}\mathcal{P}_\text{th}\cos \upphi} \right) }\right ]\tilde{\mathbb{I}}_1
					\end{align}
with $\tilde{\mathbb{I}}_1$ as 
					\begin{align}\label{it4}
						\tilde{\mathbb{I}}_1&= \int_{w_{\min}}^{w_{\max}}  w^{2k-k_1+k_2+1}\exp\left(-\frac{\bar{\beta}_{sr}w^2}{\mathcal{C}} \left(\mathcal{D}+\frac{\mathcal{P}_{\mathcal{C,D}}}{2}\right) \right)\mathbb{W}_{\frac{-p_1+k_1-k_2}{2},\frac{p_1-k_1+k_2-1}{2}}\left(\frac{\bar{\beta}_{sr}w^2 \mathcal{P}_{\mathcal{C,D}}}{\mathcal{C}} \right)dw.
					\end{align}
				Further, $\widetilde{\Psi}_3(\gamma_\textmd{A})$ is  
					\begin{align}\label{eq30}
						\widetilde{\Psi}_{3}(\gamma_\textmd{A})&=\frac{6\alpha _{sr}}{w_{er}w_{\min}\nu_{rt}\left( {h_{2}^{3}-h_{1}^{3}} \right)} \sum _{k =0}^{m_{sr}-1}  \zeta\left ({k }\right)\sum_{m=0}^{\infty }\sum_{n=0}^{m}\frac{K_{rt}^{m} \exp(-K_{rt})}{m! n!} \nonumber\\
						&\times\sum_{k_{1}=0}^{k+n}\binom{k+n}{k_{1}}\sum _{k_2 =0}^{\infty} \frac{(-1)^{k_2}}{k_2!}\frac{{ \mathcal{D}} ^{\frac{2k+n-k_1+1}{2}}}{\bar{\beta}_{sr}^{\varsigma_3}~\mathcal{C}^{k+n+k_2+1}} \left( \frac{\eta_s{\Theta(\gamma_{A})}}{\mathcal{C}\mathcal{P}_\text{th}} \right)^{n+k_2+\varsigma_3}[\tilde{\mathbb{I}}_2-\tilde{\mathbb{I}}_3-\tilde{\mathbb{I}}_4+\tilde{\mathbb{I}}_5]
					\end{align}
			 with $\tilde{\mathbb{I}}_2$, $\tilde{\mathbb{I}}_3$, $\tilde{\mathbb{I}}_4$, and $\tilde{\mathbb{I}}_5$, respectively, as
					\begin{align}\label{i5}
						\tilde{\mathbb{I}}_2&=	\int_{w_{\min}}^{w_{\max}} \left[ {h_2}^{\varsigma_2{\nu_{rt}+3}} G^{2,1}_{1,3}
						\bigg[\frac{\eta_s\Theta(\gamma_{A})h_2^{\nu_{rt}}\mathcal{D}\bar{\beta}_{sr}w^2}{\mathcal{C}^{2}\mathcal{P}_\textmd{th}} \bigg\vert{{1-\varsigma_2-\frac{3}{\nu_{rt}}}\atop{{\varsigma_3,-\varsigma_3}},-\varsigma_2-\frac{3}{\nu_{rt}}}\bigg]\right. \nonumber\\
						&\left.- {h_1}^{\varsigma_2{\nu_{rt}+3}} G^{2,1}_{1,3}
						\bigg[\frac{\eta_s\Theta(\gamma_{A})h_1^{\nu_{rt}}\mathcal{D}\bar{\beta}_{sr}w^2}{\mathcal{C}^{2}\mathcal{P}_\textmd{th}} \bigg\vert{{1-\varsigma_2-\frac{3}{\nu_{rt}}}\atop{{\varsigma_3,-\varsigma_3}},-\varsigma_2-\frac{3}{\nu_{rt}}}\bigg]\right] w^{2k+n-k_1+2}\exp\left( -\frac{\mathcal{D}\bar{\beta}_{sr}w^2}{\mathcal{C}} \right)dw, 
					\end{align}
					\begin{align}\label{i6}
						\tilde{\mathbb{I}}_3&=\int_{w_{\min}}^{w_{\max}} h_1\left[ {\left( \frac{h_1}{\cos \upphi} \right)}^{\varsigma_2{\nu_{rt}+2}}G^{2,1}_{1,3}
						\bigg[\frac{\eta_s\Theta(\gamma_{A})\mathcal{D}\bar{\beta}_{sr}w^2h_1^{\nu_{rt}}}{\mathcal{C}^{2}\mathcal{P}_\textmd{th}\cos^{\nu_{rt}} \upphi} \bigg\vert{{1-\varsigma_2-\frac{2}{\nu_{rt}}}\atop{{\varsigma_3,-\varsigma_3}},-\varsigma_2-\frac{2}{\nu_{rt}}}\bigg]\right. \nonumber\\
						&\left.- {h_1}^{\varsigma_2{\nu_{rt}+2}} G^{2,1}_{1,3}
						\bigg[\frac{\eta_s\Theta(\gamma_{A})h_1^{\nu_{rt}}\mathcal{D}\bar{\beta}_{sr}w^2}{\mathcal{C}^{2}\mathcal{P}_\textmd{th}} \bigg\vert{{1-\varsigma_2-\frac{2}{\nu_{rt}}}\atop{{\varsigma_3,-\varsigma_3}},-\varsigma_2-\frac{2}{\nu_{rt}}}\bigg]\right] w^{2k+n-k_1+2}\exp\left( -\frac{\mathcal{D}\bar{\beta}_{sr}w^2}{\mathcal{C}} \right)dw, 
					\end{align}
					\begin{align}\label{i7}
						&\tilde{\mathbb{I}}_4=\int_{w_{\min}}^{w_{\max}}\cos \upphi\!\! \left[\! {\left( \frac{h_2}{\cos \upphi} \right)}^{\varsigma_2\nu_{rt}+3}  G^{2,1}_{1,3}
						\bigg[\frac{\eta_s\Theta(\gamma_{A})\mathcal{D}\bar{\beta}_{sr}w^2h_2^{\nu_{rt}}}{\mathcal{C}^{2}\mathcal{P}_\textmd{th}\cos^{\nu_{rt}} \upphi} \bigg\vert{{1-\varsigma_2-\frac{3}{\nu_{rt}}}\atop{{\varsigma_3,-\varsigma_3}},-\varsigma_2-\frac{3}{\nu_{rt}}}\bigg]\right.\!\!- \!{\left( \frac{h_1}{\cos \upphi} \right)}^{\varsigma_2\nu_{rt}+3} \nonumber\\
						&\left.\times G^{2,1}_{1,3}
						\bigg[\frac{\eta_s\Theta(\gamma_{A})\mathcal{D}\bar{\beta}_{sr}w^2h_1^{\nu_{rt}}}{\mathcal{C}^{2}\mathcal{P}_\textmd{th}\cos^{\nu_{rt}} \upphi} \bigg\vert{{1-\varsigma_2-\frac{3}{\nu_{rt}}}\atop{{\varsigma_3,-\varsigma_3}},-\varsigma_2-\frac{3}{\nu_{rt}}}\bigg]\right] w^{2k+p_1-k_1+2}\exp\left( -\frac{\mathcal{D}\bar{\beta}_{sr}w^2}{\mathcal{C}} \right)dw,
					\end{align}
and	
					\begin{align}\label{i8}
						\tilde{\mathbb{I}}_5&=	\int_{w_{\min}}^{w_{\max}} h_2\left[ {\left( \frac{h_2}{\cos \upphi} \right)}^{\varsigma_2\nu_{rt}+2}  G^{2,1}_{1,3}
						\bigg[\frac{\eta_s\Theta(\gamma_{A})\mathcal{D}\bar{\beta}_{sr}w^2h_2^{\nu_{rt}}}{\mathcal{C}^{2}\mathcal{P}_\textmd{th}\cos^{\nu_{rt}} \upphi} \bigg\vert{{1-\varsigma_2-\frac{2}{\nu_{rt}}}\atop{{\varsigma_3,-\varsigma_3}},-\varsigma_2-\frac{2}{\nu_{rt}}}\bigg]\right. - {h_2}^{\varsigma_2\nu_{rt}+2}\nonumber\\
						&\left.\times G^{2,1}_{1,3}
						\bigg[\frac{\eta_s\Theta(\gamma_{A})\mathcal{D}\bar{\beta}_{sr}w^2h_2^{\nu_{rt}}}{\mathcal{C}^{2}\mathcal{P}_\textmd{th}} \bigg\vert{{1-\varsigma_2-\frac{2}{\nu_{rt}}}\atop{{\varsigma_3,-\varsigma_3}},-\varsigma_2-\frac{2}{\nu_{rt}}}\bigg]\right] w^{2k+p_1-k_1+2}\exp\left( -\frac{\mathcal{D}\bar{\beta}_{sr}w^2}{\mathcal{C}} \right)dw. 
					\end{align}
			\end{theorem}	
			\begin{IEEEproof}
				See	Appendix E.
			\end{IEEEproof}
			The integrals $\tilde{\mathbb{I}}_2$, $\tilde{\mathbb{I}}_3$,   $\tilde{\mathbb{I}}_4$ and $\tilde{\mathbb{I}}_5$ can be evaluated using CGQ method as described previously. 
			\subsection{With Perfect Interference Cancellation}
				The OP of the A2A network under p-IC can be evaluated similar to (\ref{eq31}) as
			\begin{align}\label{eq32m}
				\mathcal{P}^{\textmd{A2A}}_{\textmd{out}}(\gamma_\text{A}) 
				&= 1-\underbrace{\textmd{Pr}\left[\Lambda^\text{p-IC}_{T,1}  > \gamma_\text{A},\eta_s\left| g_{sr} \right|^{2}w_{sr}^{-2}\leq \mathcal{P}_\text{th} \right] }_{	\mathcal{P}_5(\gamma_\text{A})}-\underbrace{\textmd{Pr}\left[ \Lambda^\text{p-IC}_{T,2}  > \gamma_\text{A},\eta_s\left| g_{sr} \right|^{2}w_{sr}^{-2}> \mathcal{P}_\text{th} \right] }_{	\mathcal{P}_6(\gamma_\text{A})},
			\end{align}
			based on $\Lambda^\text{p-IC}_{T,1}$ and $\Lambda^\text{p-IC}_{T,2}$ as given in (\ref{n25}) and (\ref{n26}), respectively. 
			Thus, we have the following theorems.  
			\begin{theorem}	\label{theo4}
				The probability $\mathcal{P}_{5}(\gamma_\text{A})$ in (\ref{eq32m}) is given by
				\begin{equation}\label{eq35m}
					\mathcal{P}_5(\gamma_\textmd{A})=\left\{\!\begin{array}{ll}
					\overline{\Psi}_1(\gamma_\textmd{A}), &{\text{for }}\gamma_{\textmd{A}}< 1/\mu^\prime \mbox{ and } {\mathcal{P}_\text{th}}\geq\left(\frac{\mathcal{D}}{\mathcal{E}}\right){\eta_s},\\
						0, &{\text{for }}\gamma_{\textmd{A}}< 1/\mu^\prime \mbox{ and } {\mathcal{P}_\text{th}}<\left(\frac{\mathcal{D}}{\mathcal{E}}\right){\eta_s},
					\end{array}\right.
				\end{equation}
			
					\end{theorem}	
				\begin{IEEEproof} 	
			The evaluation of $\mathcal{P}_5(\gamma_\textmd{A})$ follows a similar procedure to that used for deriving $\mathcal{P}_3(\gamma_\textmd{A})$, with the parameter $\mathcal{C}$ replaced by $\mathcal{E}$.
				\end{IEEEproof}
		   
				\begin{theorem}	\label{theo3}
				The probability $\mathcal{P}_{6}(\gamma_\text{A})$ in (\ref{eq32m}) is given by,
				\begin{equation}\label{eq35b}
					\mathcal{P}_6(\gamma_\textmd{A})=\left\{\!\begin{array}{ll}
						\overline{\Psi}_2(\gamma_\textmd{A}), &{\text{for }}\gamma_{\textmd{A}}< 1/\mu^\prime \mbox{ and } {\mathcal{P}_\text{th}}\geq\left(\frac{\mathcal{D}}{\mathcal{E}}\right){\eta_s},\\
							\overline{\Psi}_3(\gamma_\textmd{A}), &{\text{for }}\gamma_{\textmd{A}}< 1/\mu^\prime \mbox{ and } {\mathcal{P}_\text{th}}<\left(\frac{\mathcal{D}}{\mathcal{E}}\right){\eta_s},
					
					\end{array}\right.
				\end{equation}
					\end{theorem}	
				\begin{IEEEproof}
				 The exact  expressions of $\mathcal{P}_6(\gamma_\textmd{A})$ for p-IC scenario can be obtained directly from $\mathcal{P}_4(\gamma_\textmd{A})$  by replacing the parameter $\mathcal{C}$ by $\mathcal{E}$.
				\end{IEEEproof}
			It is worth mentioning that when the aerial node can decode the signal, unlike in the preceding scenario, the limitations on the spectrum sharing factor $\mu$ is relaxed. Thus, it can be concluded that the performance of the secondary system can be greatly enhanced when the aerial node has the capability of correctly decoding the signal, as we will see further through the numerical results.
			\section{Average System Throughput}
			The average system throughput serves as a vital performance metric for assessing the spectrum utilization of the OSAGIN system under consideration. Given that the average transmission rates for S2G and A2A networks are $r_{s}$ and $r_{a}$ bits/sec/Hz, respectively, and $(1-\rho)\mathcal{T}/2$ is the effective transmission time, the average system throughput can be expressed according to \cite{18} as
			\begin{align}\label{thp} 
				\mathcal {S}_{\text {avg}} &={\left(1-\rho \right) }\mathcal{T}/2[r_{s}\left(1-	\mathcal{P}^{\textmd{S2G}}_{\textmd{out}}(\gamma_\textmd{S}) \right) 
				+r_{a}\left(1-	\mathcal{P}^{\textmd{A2A}}_{\textmd{out}}(\gamma_\textmd{A})  \right) ].  
			\end{align}
		In (\ref{thp}), the first and second terms represent individually the average throughput of S2G and A2A networks, respectively. Note that the $\mathcal {S}_{\text {avg}}$ is a function of parameter $\mu$, where $\mu=1$ allocates the zero transmit power for the A2A communications. Under this special condition, the A2A communication ceases and the considered OSAGIN operates as a conventional/standalone S2G network without spectrum sharing. 
					
				\section{Numerical Results}\label{num}
				In this section, we present numerical and simulation results to illustrate the impact of key system and channel parameters on the performance of considered OSAGIN. For this purpose, we set SR fading parameters $\left\{ m_{sr}, \flat_{sr}, \Omega_{sr} \right\}$ corresponding to the satellite link as $\left\{ 2, 0.063, 0.0005 \right\} $ and $\left\{ 5, 0.251, 0.279 \right\} $ to render heavy and light shadowing, respectively. We set other satellite parameters as $\mathbb{T}=300$ K, $\mathcal{W}=15$ MHz, $\mathcal{K_B}=1.38\times10^{-23}$J/K, $\xi=2~\textmd{dB}$, $\lambda=0.15$ m, $\vartheta_{sr}=4.8$ dB, $\vartheta_{s}=53.45$ dB, $\theta_{sr}=0.8^{\circ}$, and $\theta_{sr 3\textmd{dB}}=0.3^{\circ}$. The fading parameters pertaining to the terrestrial and aerial links are set as $m_{rd_i}=m_{rd}=2$, $\forall i$ (under i.i.d. channels), $K_{rt} = 1$,  $\sigma ^{2}_{d}=\sigma ^{2}_{t}=\sigma ^{2}_{rb}=\sigma ^{2}_{r}= -50~ \textmd{dBm}$. We further set $w_E= 6371$ km,  $w_{\min}=400~\textmd{km} $,  $h_0 =800~\textmd{m}$,   $l= 250~\textmd{m}$, $h_1 =400~\textmd{m}$, $h_2 =500~\textmd{m}$,   $\nu_{rd_i}=\nu_{rt} =2 $, $\upphi=\frac{\pi}{12}~\textmd{ rad}$,  $l'=200$ m, $\mathcal{T}$ = $1$ sec, $\chi =0.6$, $\mu=0.7$,  $\rho=0.4$, $\varepsilon = 0.4$, and $\gamma_{\textmd{S}}=\gamma_{\textmd{A}}=5~ \textmd{dBm}$ unless stated otherwise in the specific figures.
				
				\begin{figure}[!t]
					\centering	
					\includegraphics[width=3.0in]{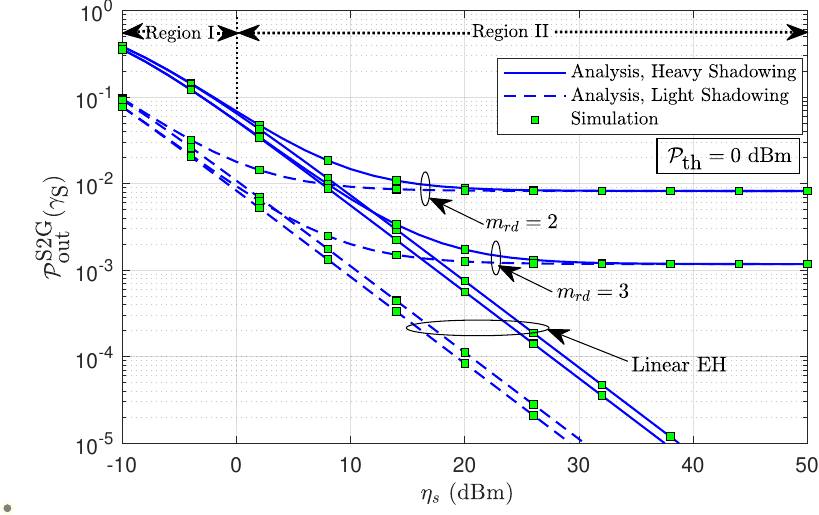}
					\caption{OP of the S2G network versus $\eta_{s}$ for different fading conditions.}
					\label{fig4}
				\end{figure}
				In Fig. \ref{fig4}, we plot the OP curves of the S2G network against $\eta_s$ under different fading conditions of satellite and GU links. Since a piecewise non-linear EH model has been adopted at the ATx, the OP of the S2G network is assessed in Regions I and II. Here, Region I corresponds to the linear EH mode while Region II refers to the saturation mode for the operation of non-linear energy harvester. We observe that in Region I, the OP of the S2G network decreases sharply as $\eta_s$ increases for given parameter settings. For instance, when $m_{rd}$ is fixed (say $2$ or $3$), the OP under average and heavy shadowing over satellite link decreases rapidly with $\eta_s$ in Region I. On the other hand, in Region II, we observe that for a given $m_{rd}$, the OP curves under both the average and heavy shadowing merge into a floor. The outage floor appears in Region II because the peak harvested power is limited by the saturation threshold power $\mathcal{P}_\textmd{th}$. Here, the OP curves under average and heavy shadowing become indistinguishable that limits the use of a better satellite link condition in Region II. However, when $m_{rd}$ takes on a higher value (say $3$), we observe a downward shift in the floor's position in Region II as GU link channel becomes better. For comparison purpose, we further plot the OP curves of the S2G network with linear EH under the same parameter settings. Here, we observe that for a fixed value of $m_{rd}$, the OP of S2G network is significantly better when the satellite link's channel is better. Here, unlike non-linear EH scenario, two OP curves corresponding to average and heavy shadowing do not merge over the entire SNR regime. Therefore, the satellite link channel contributes significantly in achieving performance gain for the S2G network. In subsequent figures, we produce the plots by considering heavy shadowing exclusively for the satellite link unless otherwise stated.     
							
				\begin{figure}[!t]
					\centering	
					\includegraphics[width=3.0in]{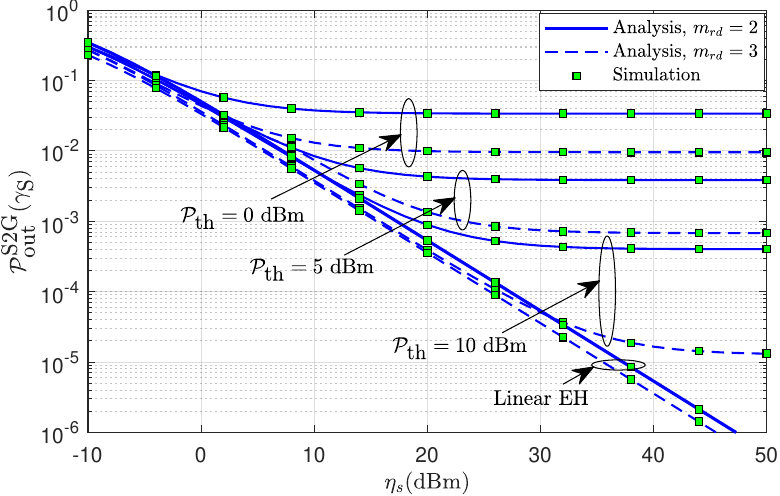}
					\caption{OP of the S2G network versus $\eta_{s}$ for different values of ${\mathcal{P}_\text{th}}$.}
					\label{fig8}
				\end{figure}
				In Fig. \ref{fig8}, we plot the OP curves of the S2G network against $\eta_s$ for variable saturation threshold power $\mathcal{P}_\textmd{th}$. We observe that for a given value of $m_{rd}$ (i.e., $2$ or $3$), the OP curves under both the average and heavy shadowing over the satellite link decreases linearly up to a certain value of $\eta_s$. The linear relationship of OP curves with $\eta_s$ depends on the value of $\mathcal{P}_\textmd{th}$. For instance, when $\mathcal{P}_\textmd{th}$ increases from $0$ dBm to $5$ (or $10$) dBm, the outage floor appears at a higher level of $\eta_s$. This results in a downward shift of the level of floor in saturation region that leads to a better outage performance of the S2G network at higher value of $\mathcal{P}_\textmd{th}$. Alternatively, the region for linear operation extends as the higher value of $\mathcal{P}_\textmd{th}$ results in an increased peak harvested power in saturation mode. Further, the OP curves with linear EH do not exhibit any floor in the high SNR regime.      
				
				\begin{figure}[!t]
					\centering	
					\includegraphics[width=3.0in]{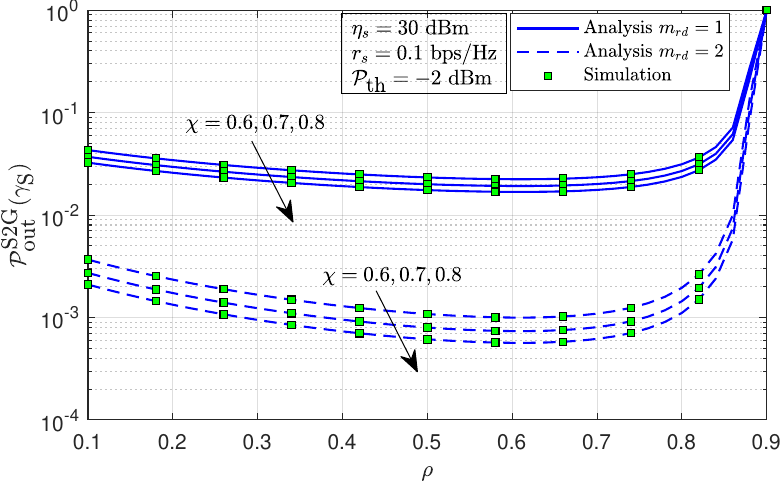}
					\caption{OP of the S2G network versus $\rho$ for different values of $\chi$.}
					\label{fig5}
				\end{figure}
				Fig. \ref{fig5} illustrates the impact of time fraction $\rho$ and energy conversion efficiency $\chi$. We see that for given values of $m_{rd}$ and $\chi$, the OP curves of S2G network take on a convex shape yielding the minimum at certain value of $\rho$ (say $\rho^\ast$). It implies that there exists an optimal value of $\rho$ (i.e., $\rho^\ast\approx0.61$) that minimizes the OP of S2G network. Further, we observe that when $\chi$ increases (e.g., from $0.6$ to $0.7$ (or $0.8$)), the OP of S2G network improves due to efficient energy harvesting at ATx for WPT. However, when $\rho\leq\rho^\ast$ and $\rho>\rho^\ast$, the OP of S2G network deteriorates. This occurs because, while the former case shortens the EH phase (i.e., leading to poorer energy harvesting), the latter case abbreviates the IP phase (i.e., limiting the information transmission). Apparently, the OP improves when $m_{rd}$ increases as the system exhibits enhanced link reliability under relatively less severe GU link's fading conditions.

				\begin{figure}[!t]
					\centering	
					\includegraphics[width=3.0in]{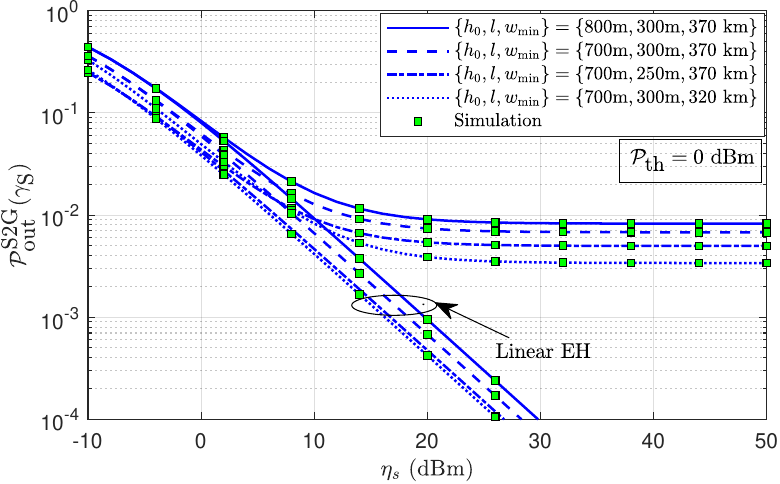}
					\caption{OP of the S2G network versus $\eta_{s}$ for different distance parameters.}
					\label{fig7}
				\end{figure}
				Fig. \ref{fig7} shows the OP curves against $\eta_s$ for variable distance parameters, i.e., $h_0$, $l$, and $w_{\min}$. Considering the plot for parameters $\{h_0, l, w_{\min}\}=\{700~\text{m}, 300~\text{m}, 370~\text{km}\}$ as a reference, we compare it successively with those obtained for $\{800~\text{m}, 300~\text{m}, 370~\text{km}\}$, $\{700~\text{m}, 250~\text{m}, 370~\text{km}\}$, and  $\{700~\text{m}, 300~\text{m}, 320~\text{km}\}$. We observe that in the first case, the OP of the S2G network degrades as the distance $h_0$ increases. This results in an increased path loss over the GU link. On the contrary, the OP of S2G network improves in the second and third cases as the distances $l$ and $w_{\min}$ reduces due to reduced path loss over GU and satellite links, respectively. We have further plotted OP curves with linear EH for comparison purpose. Here, the OP curves do not make floor at high SNR.       
				
				\begin{figure}[!t]
					\centering	
					\includegraphics[width=3.0in]{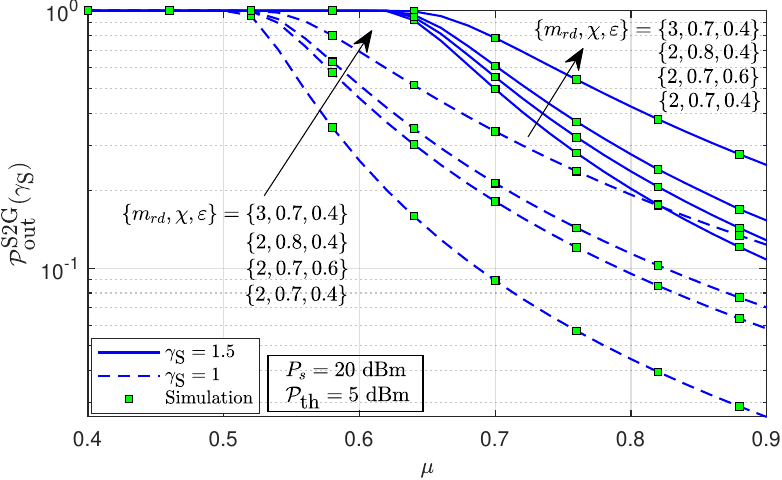}
					\caption{OP of the S2G network versus $\mu$  for different values of $\gamma_\textmd{S}$.}
					\label{fig9}
				\end{figure}
				Fig. \ref{fig9} presents the OP curves for S2G network against $\mu$ for different values of $\gamma_\textmd{S}$ and other parameters.  According to Theorems $1$ and $2$, the non-unity OP of S2G network is only valid for the condition $\gamma_\textmd{S}<\mu^\prime$. It implies that the OP remain unity until $\mu$ attains a critical value $\mu^\ast$ (or higher) for given value of $\gamma_\textmd{S}$. The value $\mu\geq\mu^\ast=\gamma_\textmd{S}/(\gamma_\textmd{S}+1)$ indicates the feasible operational range for the considered system. For instance, when $\gamma_\textmd{S}=1$, the OP remains unity till $\mu$ attains the value $\mu^\ast=0.5$ (or $0.6$). In fact, the condition $\gamma_\textmd{S}<\mu^\prime$ yields the minimum possible value of $\mu^\ast$ as $0.5$. This infers that at least $50$\% of the transmit power must be reserved for the reliable signal transmission (minimum targeted QoS) of the S2G network. We have further plotted the OP curves by varying the parameters $\{m_{rd}, \chi, \varepsilon\}$. On comparing the curves under different settings of these parameters, we observe that for a given value of $\gamma_\textmd{S}$, the value of $\mu^\ast$ remains unaltered. However, the OP curves varies according to the parameter values when $\mu$ exceeds $\mu^\ast$. Apparently, the outage  performance of the S2G network is better for a lower target threshold $\gamma_\textmd{S}$ because it results in a relatively lower target SNR.

				\begin{figure}[!t]
					\centering	
					\includegraphics[width=3.0in]{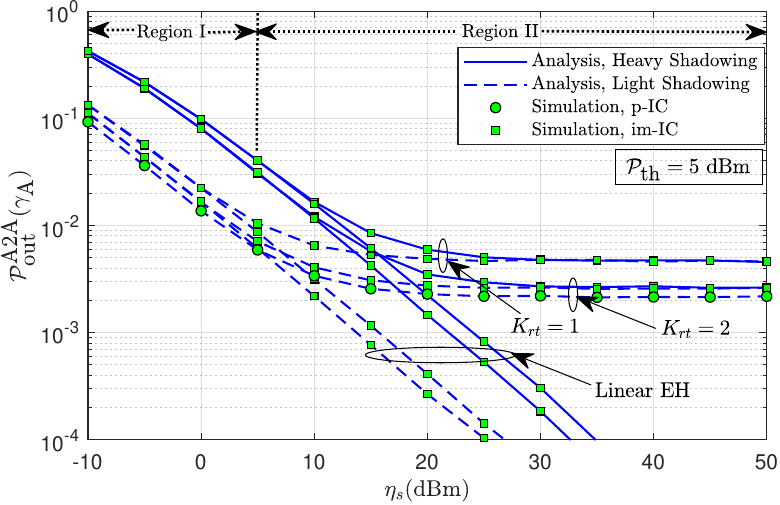}
					\caption{OP of the A2A network versus $\eta_{s}$ for different fading conditions.}
					\label{fig10}
				\end{figure}
				In Fig. \ref{fig10}, we show the OP curves of the A2A network versus $\eta_s$ under different fading scenarios pertaining to the satellite and ARx links. Similar to as in Fig. \ref{fig4}, the outage performance of the A2A network is assessed in Regions I and II. It is observed that in Region I, the OP of the A2A network falls off rapidly as $\eta_s$ increases for given parameter settings. In particular,  when $K_{rt}$ is kept constant (e.g., $1$ or $2$), the OP decreases swiftly with $\eta_s$ in Region I under both average and heavy shadowing over the satellite link. Conversely, within Region II, we notice that regardless of the specified value of $K_{rt}$, the OP curves end-up in a floor under both types of shadowing conditions. Here also, the occurrence of the floor in Region II is due to the saturation of harvested power beyond $\mathcal{P}_\textmd{th}$. Further, in Region II, the OP curves under both the shadowing conditions overlap, restricting the performance gain via better satellite link. Nevertheless, when $K_{rt}$ assumes a higher value (e.g., $2$), the outage floor shifts downwards within Region II as the channel corresponding to the ARx link becomes better. Additionally, we have the OP curves of the A2A networks with linear EH under identical parameter configurations. Here, we observe that for a fixed value of $K_{rt}$, the OP of the A2A network improves significantly when the satellite link's channel quality is better. Also, no floor occurs at the high SNR. Therefore, the satellite link channel can be exploited to harness performance gain for the A2A network. Besides, we plotted an exclusive OP curve for p-IC case considering light shadowing and $K_{rt}=2$. We find that the A2A network achieves better performance under p-IC as compared to that under im-IC.

				\begin{figure}[!t]
					\centering	
					\includegraphics[width=3.0in]{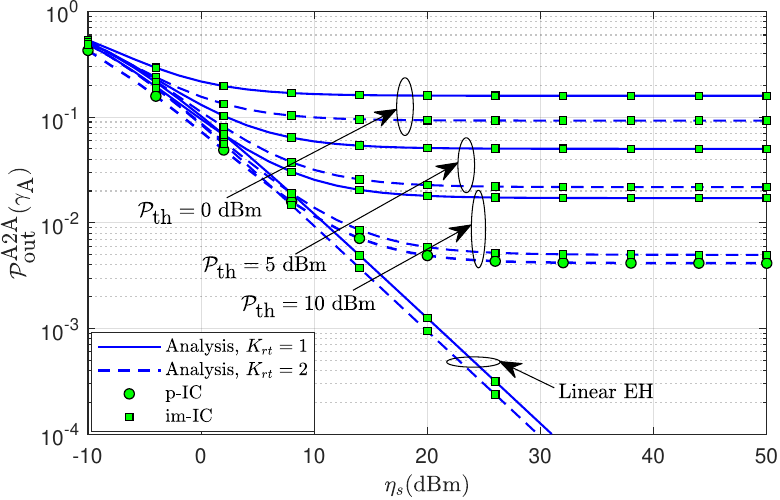}
					\caption{OP of the A2A network versus $\eta_{s}$ for different values ${\mathcal{P}_\text{th}}$.}
					\label{fig11}
				\end{figure} 
				In Fig. \ref{fig11}, the OP curves of the A2A network are plotted against $\eta_s$ for varying saturation threshold power $\mathcal{P}_\textmd{th}$. We observe that for a given value of $K_{rt}$ (i.e., $1$ or $2$), the OP curves  decrease linearly up to a certain value of $\eta_s$ depending on the value of $\mathcal{P}_\textmd{th}$. For instance, when $\mathcal{P}_\textmd{th}$ increases from $0~ \textmd{dBm}$  to $5$ (or $10$) $ \textmd{dBm}$, the outage floor appears at a relatively higher level of $\eta_s$. Alternatively, there is a downward shift in the level of the floor that yields a better outage performance of the A2A network when $\mathcal{P}_\textmd{th}$ at a higher value. Further, the OP curves with linear EH do not exhibit any floor in the high SNR regime. Additionally, we examined the OP of the A2A network under p-IC for a representative case when ${\mathcal{P}_\text{th}}=10$ dBm. With p-IC, the OP of the A2A network is better than that of its im-IC counterpart.    
				
\begin{figure}[t]
	\centering	
	\includegraphics[width=3.0in]{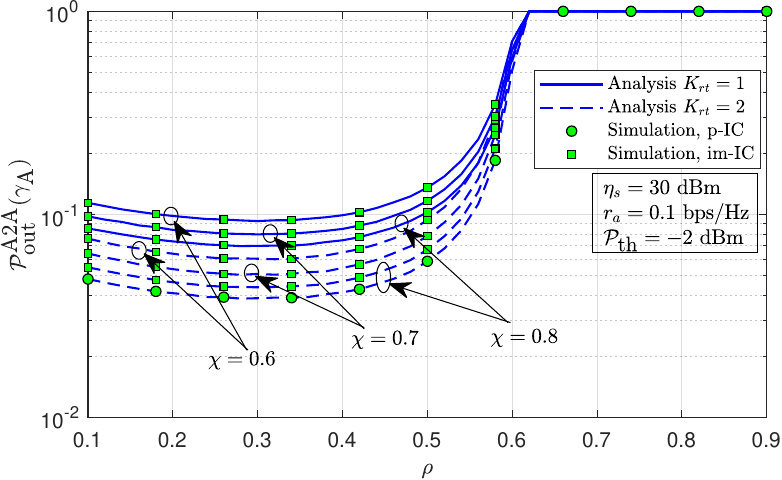}
	\caption{OP of the A2A network versus $\rho$ for different value of $\chi$.}
	\label{fig12}
\end{figure}
Fig. \ref{fig12} assesses the impact of time fraction $\rho$ and energy conversion efficiency $\chi$ on the OP of A2A network. We observe that the OP curves of the A2A network exhibit a convex shape with respect to $\rho$. This implies the existence of an optimal value of $\rho$ (i.e., $\rho^\ast\approx 0.3$) minimizing the OP of A2A network like S2G network. However, the values of $\rho$ individually minimizing the OP of A2A and S2G networks are not necessarily the same. This is due to the fact that the A2A network has dissimilar network configuration (i.e., geometry and Rician fading) as compared to that of the S2G network. On further increasing the $\rho$ beyond $\rho^\ast$, the OP of the A2A network quickly bottlenecks and reach unity at $\rho\geq 0.62$ under the considered settings.  Apparently, when $\chi$ increases (e.g., from $0.6$ to $0.7$ (or $0.8$)), the improvement in OP of A2A network is caused by the efficient energy harvesting at ATx. As expected, the OP improves with increasing  $K_{rt}$, as the system exhibits an enhanced link reliability under relatively less severe ARx link's fading conditions. Moreover, the A2A network with p-IC outperforms that with im-IC.
				
\begin{figure}[!t]
	\centering	
	\includegraphics[width=3.0in]{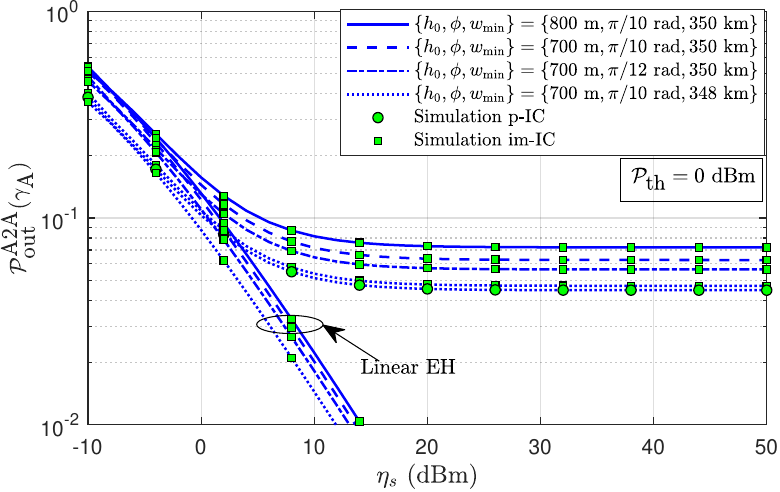}
	\caption{OP of the A2A network versus $\eta_{s}$ for different distance parameters.}
	\label{fig14}
\end{figure} 
Fig. \ref{fig14} shows the OP curves against $\eta_s$ for various distance related parameters, i.e., $h_0$, $\upphi$, and $w_{\min}$. Using the plot for the parameters $\left\{h_0, \upphi, w_{\min}\right\}=\left\{700~\text{m}, \pi/10, 350~\text{km}\right\}$ as a benchmark, we compare it sequentially with those obtained for $\left\{800~\text{m}, \pi/10, 350~\text{km}\right\}$, $\left\{700~\text{m}, \pi/12, 350~\text{km}\right\}$, and $\left\{700~\text{m}, \pi/10, 348~\text{km}\right\}$. We observe that in the first case, the OP of the A2A network deteriorates as the distance $h_0$ increases, leading to increased path loss over the aerial link. In the second case, the OP of A2A network improves as the half-beamwidth angle $\upphi$ decreases. This is because a lower value of $\upphi$ is linked to a smaller deployment region for the ARx, which eventually reduces the path-loss. Likewise, in the third case, the OP of the A2A network improves as the distance $w_{\min}$ decreases. Nevertheless, the OP curves with linear EH do not exhibit a floor at high SNR. In addition, the p-IC at ARx yields slightly better performance with respect to im-IC.

	\begin{figure}[!t]
		\centering	
		\includegraphics[width=3.0in]{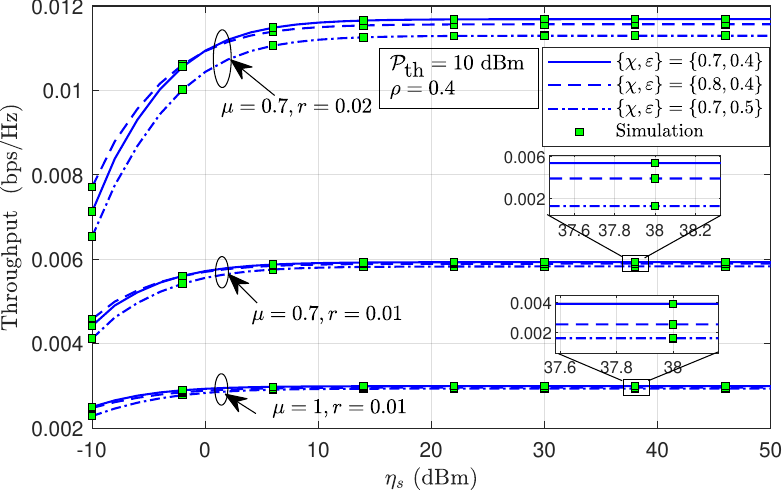}
		\caption{System throughput versus $\eta_{s}$ for different values of $\mu$ and $r$.}
		\label{fig15}
	\end{figure}
	Fig. \ref{fig15} illustrates the average system throughput curves against $\eta_{s}$ for different values of $\mu$ and $r$. Here, we set the parameters $r=r_s=r_a$, where $r$ takes two values $0.01$ and $0.02$ bits/s/Hz, when $\mu\neq1$. We further consider the curves plotted for $r=0.02$ bits/s/Hz and $\mu=1$ as reference. Note that the case with $\mu=1$ represents the throughput curves for a standalone S2G network without spectrum sharing, i.e., not supporting A2A communications. We also vary the parameters $\{\chi,\varepsilon\}$ for given values of $\mu$ and $r$. Let us first compare the set of curves for $\mu=1, r=0.02$ and $\mu=0.7, r=0.02$ bits/s/Hz. Here, we can see that the average system throughput under $\mu=0.7, r=0.02$ bits/s/Hz is remarkably better than that under $\mu=1, r=0.02$ bits/s/Hz. This implies that for a given rate, i.e., $r=0.02$ bits/s/Hz, the considered OSAGIN with spectrum sharing (i.e., $\mu\neq1$) offers very large throughput (or spectral efficiency) than that of a standalone S2G network without spectrum sharing (i.e., $\mu=1$). When $\mu\neq1$, both the S2G and A2A communications take place successfully that result in a combined throughput pertaining to each of these networks. Further, we compare the throughput curves for $\mu=1, r=0.02$ bits/s/Hz and $\mu=0.7, r=0.01$ bits/s/Hz. Even, when $r=0.01$ bits/s/Hz (for $\mu=0.7$), the average system throughput is significantly larger than that of standalone S2G network (i.e., $\mu=1$) with $r=0.02$ bits/s/Hz. Furthermore, we see that for fixed values of $\mu$ and $r$ (say $\mu=0.7$ and $r=0.02$), the system throughput changes marginally with respect to the parameters $\{\chi,\varepsilon\}$ at high SNR.    
	
\begin{figure}[!t]
	\centering	
	\includegraphics[width=3.0in]{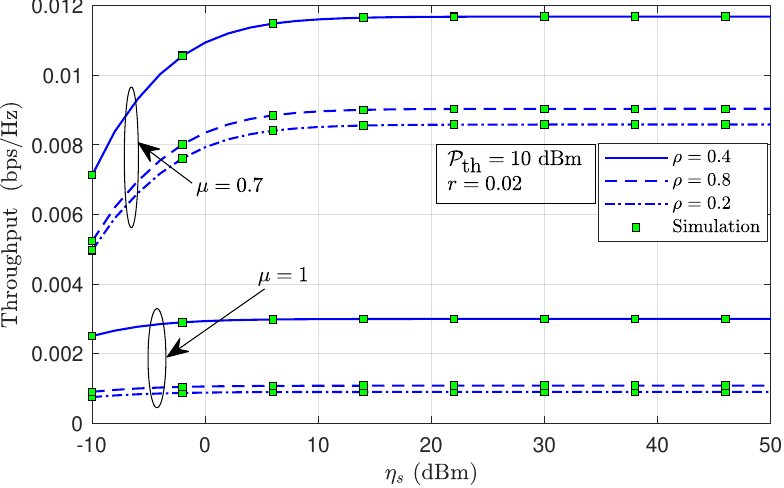}
	\caption{System throughput versus $\eta_{s}$ for different values of $\rho$.}
	\label{fig16}
\end{figure}	
Fig. \ref{fig16} shows the average system throughput curves against $\eta_{s}$ for different values of $\rho$. Here also, we consider the case of $\mu=1$ as reference which represents pure S2G network without spectrum sharing for A2A communications. Here, we can see that for a given value of $\rho$ (i.e., $0.2$, $0.4$, or $0.8$), the average system throughput of the considered OSAGIN is significantly larger with $\mu=0.7$ as compared to that under $\mu=1$. As revealed previously, this is due to the combined throughput of successful S2G and A2A communications under $\mu\neq1$. Clearly, the considered OSAGIN with spectrum sharing offers a higher spectral efficiency as compared to that of an only S2G network. Notably, $\rho$ acts as a determining factor for the durations of EH and IP phases as well as a scale factor in the throughput expression in (\ref{thp}). Hence, its value critically affects the system throughput. For instance, when $\rho$ changes from $0.4$ to $0.8$ (or $0.2$), the system throughput is found to be decreased. This is due to the fact that when $\rho^\ast$ is closer to $0$ or $1$, the duration of the corresponding EH phase or the IP phase decreases that results in a bottleneck system performance. 

\section{Conclusion}
This paper has explored the integration of LEO satellites with ground-based and/or aerial networks, which is deemed essential for facilitating future 6G communications. Specifically, we assessed the performance of OSAGIN utilizing a non-linear hybrid SWIPT-based spectrum sharing system to enable S2G and A2A communications over the same licensed spectrum. The considered system model differs from the existing studies as it takes into account an aerial transmitter ATx equipped with SWIPT capabilities to extract energy from the satellite’s signal. Additionally, we studied the randomness of the satellite, ARx, and ground user locations, along with appropriate channel fading in S2G and A2A networks, to derive corresponding OP expressions. Further, we assumed both p-IC and im-IC at ARx and demonstrated that the performance of the aerial network can be significantly improved when p-IC is possible. Numerical and simulation results validate the accuracy of the derived analytical expressions. The influence of key design parameters is illuminated on the performance of the system under consideration. Numerical results emphasized the importance of integrating a SWIPT-enabled aerial network for self-reliant energy-efficient and spectral-efficient communications. We also demonstrated that a judicious choice of time-splitting factor and spectrum sharing factor is essential in order to maximize the system throughput. Moreover, the numerical analysis depicted that the OP of both S2G and A2A networks improves when $\chi$ increases and/or the link distances decreases. 
	
\begin{appendices}
		\section{Proof of  Theorem 1}
					
						To calculate the pdf of $w_{sr}$, the $3$D geometry for the instantaneous location of LEO satellite $S$ over a spherical surface in space with reference to $R$ and the Earth's center $E$ is shown in the Fig. \ref{leo}. In this figure, we have
						\begin{equation}\label{d1}
							w_{qe}=w_{\min}-w_{h}+h_{0}+w_{e}.   
						\end{equation}
						By applying the Pythagorean theorem in $\triangle$RSQ and $\triangle$OSQ, we obtain
						\begin{equation}\label{d4}
							w_{sr}^{2}=w_{so}^{2}-h_{0}^{2}-2h_{0}\left( w_{\min} -w_{h}\right).    
						\end{equation}
						Likewise, by applying the Pythagorean theorem in $\triangle$ESQ and $\triangle$OSQ, we obtain
						\begin{equation}\label{d7}
							w_{so}^2  =w_{se}^2+w_{e}^2-2w_{qe}w_{e}.
						\end{equation}
						Now, on substituting first (\ref{d7}) in (\ref{d4}) and then, simplifying the result using (\ref{d1}), we get  
						\begin{align}\label{d8}
							w_{sr}^2 
							&= w_{se}^2-w_{er}\left(w_{er} -2w_{\min} +2w_h\right).   
						\end{align}
						The total surface area of the spherical region above a horizontal plane coinciding the point $Q$ is given by $ \mathcal{S}=2\pi (w_{er}+w_{\min})w_{h}$. Thus, the CDF of $\mathcal{S}$ can be determined as
						\begin{equation}\label{d9}
							F_{\mathcal{S}}(\omega)=\textmd{Pr}(\mathcal{S}\leq \omega) =\frac{\omega}{2\pi w_{\min}(w_{er}+w_{\min})}.
						\end{equation}
						Further, the CDF of $w_{sr}$ can be computed as
						\begin{align} \label{d55}
							F_{w_{sr}}(w)&=\textmd{Pr}(w_{sr}\leq w)=\textmd{Pr}(w_{sr}^{2}\leq w^{2}).
						\end{align}
						In (\ref{d55}), we first invoke $w^2_{sr}$ as given in (\ref{d8}), and then representing $w_h=\frac{\mathcal{S}}{2\pi (w_{er}+w_{\min})}$, it can be re-expressed as  
							\begin{align}\label{d54}
								F_{w_{sr}}(w)&=\textmd{Pr}\left[ \mathcal{S}\leq \frac{\pi(w_{er}+w_{\min})(w^2- w_{se}^{2}+w_{er}^2+2w_{er}w_{\min})}{w_{er}} \right].
							\end{align}
						Finally, based on (\ref{d9}), the required CDF in (\ref{d54}) can be obtained as
						\begin{align} \label{d11}
							F_{w_{sr}}(w)&=\frac{w^2- w_{se}^{2}+w_{er}^2+2  w_{\min} w_{er}}{2  w_{\min} w_{er}}.
						\end{align}
						Now, the PDF $f_{w_{sr}}(w)$ in Lemma \ref{lem} can be determined by taking the derivative of $F_{w_{sr}}(w)$ as given in (\ref{d11}).
						
						\section{Proof of Theorem 2}
						On substituting $\Lambda_{D_i,1}$ as given by (\ref{eq23}) in (\ref{27}), we can evaluate $\mathcal{P}_1(\gamma_\textmd{S})$ as 
						\begin{align}\label{31}
							\mathcal{P}_1(\gamma_\textmd{S})&=\textmd{Pr}\left[Y_i  > \frac{\sigma^2_{d_i}\gamma_\textmd{S}w^{\nu_{rd_i}}_{rd_i}}{\mathcal{A}Xw_{sr}^{-2}-\mathcal{B}},X\leq \frac{\mathcal{P}_\text{th}w_{sr}^{2}}{\eta_s} \right], 
						\end{align}
						where  $X\triangleq\left| g_{sr} \right|^{2}$ and $Y_i\triangleq\left| g_{rd_i} \right|^{2}$. The solution of (\ref{31}) requires two conditions, i.e., $\mathcal{A}>0$ and $\mathcal{A}\leq0$ that imply $\gamma_{\textmd{S}}<\mu^\prime$ and $\gamma_{\textmd{S}}\geq\mu^\prime$, respectively. Here, the conditions $\gamma_{\textmd{S}}\geq\mu^\prime$ and $\mathcal{P}_\text{th}<\left(\frac{\mathcal{B}}{\mathcal{A}}\right){\eta_s}$ lead to $\mathcal{P}_1(\gamma_\textmd{S})=0$. Further, for $\gamma_{\textmd{S}}< \mu^\prime$ and $\mathcal{P}_\text{th}\geq\left(\frac{\mathcal{B}}{\mathcal{A}}\right){\eta_s}$, we determine $\mathcal{P}_1(\gamma_\textmd{S})= \Psi_1(\gamma_{\textmd{S}})$ as
						\begin{align}\label{61}
							\Psi_{\textmd{1}} (\gamma_\textmd{S})
							&=\int_{h_0}^{\sqrt{h^2_0+l^2}}f_{w_{rd_i}}(v)\int_{w_{\min}}^{w_{\max}}f_{w_{sr}}(w)\int_{\frac{\mathcal{B}w^{2}}{\mathcal{A}}}^{\frac{\mathcal{P}_\text{th}w^{2}}{\eta_s}}f_X(x)\int_{\frac{\sigma^2_{d_i}\gamma_{\textmd{S}}v^{\nu_{rd_i}}}{\mathcal{A}xw^{-2}-\mathcal{B}}}^{\infty}f_{Y_i}(y)dydxdwdv.
						\end{align} 
						Integrating (\ref{61}) w.r.t. $y$ using the pdf of $Y_i$ in (\ref{eq4}) based on \cite[eq. 3.351]{grad} and representing the upper incomplete gamma function $\Gamma(\cdot,\cdot)$ in series form \cite[eq. 8.37]{grad}, we get
						\begin{align}\label{39i}
							\Psi_{\textmd{1}} (\gamma_\textmd{S})
							&= \int_{h_0}^{\sqrt{h^2_0+l^2}}f_{w_{rd_i}}(v)\int_{w_{\min}}^{w_{\max}}f_{w_{sr}}(w)\nonumber\\
							&\times\int_{\frac{\mathcal{B}w^{2}}{\mathcal{A}}}^{\frac{\mathcal{P}_\text{th}w^{2}}{\eta_s}}f_X(x)\sum_{n=0}^{m_{rd_i}-1}\frac{1}{n!}\left(\frac{m_{rd_i}\sigma^2_{d}\gamma_\textmd{S}v^{\nu_{rd_i}}}{\mathcal{A}xw^{-2}-\mathcal{B}}\right)^n \exp\left(-\frac{m_{rd_i}\sigma^2_{d}\gamma_\textmd{S}v^{\nu_{rd_i}}}{\mathcal{A}xw^{-2}-\mathcal{B}}\right)dxdwdv.
						\end{align}
						To solve (\ref{39i}), we invoke a variable change $z=\mathcal{A}xw^{-2}-\mathcal{B}$ and substitute the pdfs of $X$ and $w$. Eventually, we employ the Taylor series for the term $\exp\left(-\frac{\bar{\beta}_{sr}zw^2}{\mathcal{A}}\right)$ and the identity  $\exp\left(-\frac{m_{rd_i}\sigma^2_{d}\gamma_\textmd{S}v^{\nu_{rd_i}}}{z}\right) = G^{0,1}_{1,0}
						\bigg[\frac{z}{m_{rd_i}\sigma^2_{d}\gamma_\textmd{S}v^{\nu_{rd_i}}} \bigg\vert{{1}\atop{{-}}}\bigg]$ to write 
							\begin{align}\label{40}
								&\Psi_{\textmd{1}}(\gamma_\textmd{S})
								=\alpha _{sr} \sum _{k =0}^{m_{sr}-1}\zeta\left ({k }\right)\sum_{k_1=0}^{k}\binom{k}{k_1}\sum_{k_2=0}^{\infty }\frac{(-1)^{k_2}\bar{\beta}_{sr}^{k_2}\mathcal{B}^{k-k_1}}{k_2!}\int_{h_0}^{\sqrt{h^2_0+l^2}}f_{w_{rd_i}}(v)\sum_{n=0}^{m_{rd_i}-1}\frac{\left( m_{rd_i}\sigma^2_{d_i}\gamma_\textmd{S}v^{\nu_{rd_i}}\right)^n}{n!}\nonumber\\
								&\times\!\int_{w_{\min}}^{w_{\max}}\frac{w}{w_{er}w_{\min}}\left( \frac{w^2}{\mathcal{A}} \right)^{k+k_2+1}\exp\left(-\frac{\bar{\beta}_{sr}\mathcal{B}w^2}{\mathcal{A}}\right)\int_{0}^{\mathcal{P}_{\mathcal{A}, \mathcal{B}}}\!\!z^{k_1+k_2-n} G^{0,1}_{1,0}
								\bigg[\frac{z}{m_{rd_i}\sigma^2_{d}\gamma_\textmd{S}v^{\nu_{rd_i}}} \bigg\vert{{1}\atop{{-}}}\bigg]dzdwdv.
							\end{align}  
						Now, we perform successively the integrations w.r.t. $z$ and $w$ by making use of \newline\cite[eq. 07.34.21.0003.01]{r121} and \cite[eq. 2.33.10]{grad}, respectively, to obtain 
							\begin{align}\label{63}
								\Psi_{\textmd{1}}(\gamma_\textmd{S})
								&=\frac{\alpha _{sr}\mathcal{A}}{2w_{er}w_{\min}} \sum _{k =0}^{m_{sr}-1}\zeta\left ({k }\right)\sum_{k_1=0}^{k}\binom{k}{k_1}\sum_{k_2=0}^{\infty }\frac{(-1)^{k_2}\mathcal{B}^{-k_1-k_2-2}}{\bar{\beta}_{sr}^{k+2}k_2!}\nonumber\\
								&\times\Delta_\Gamma\left( k+k_2+2,\frac{\bar{\beta}_{sr}\mathcal{B}w_{\min}^{2}}{\mathcal{A}},\frac{\bar{\beta}_{sr}\mathcal{B}w_{\max}^{2}}{\mathcal{A}} \right) \sum_{n=0}^{m_{rd_i}-1}{\mathcal{P}_{\mathcal{A}, \mathcal{B}}}^{\varsigma_1} \nonumber\\
								&\times\!\int_{h_0}^{\sqrt{h^2_0+l^2}}\frac{\left( m_{rd_i}\sigma^2_{d_i}\gamma_\textmd{S}v^{\nu_{rd_i}}\right)^n}{n!}G^{0,2}_{2,1}
								\bigg[\frac{{\mathcal{P}_{\mathcal{A}, \mathcal{B}}}}{m_{rd_i}\sigma^2_{d}\gamma_\textmd{S}v^{\nu_{rd_i}}} \bigg\vert{{-k_1-k_2+n,1}\atop{{-k_1-k_2+n-1}}}\bigg]f_{w_{rd_i}}(v)dv.
							\end{align}  
						Finally, on integrating (\ref{63}) w.r.t. $v$ using \cite[eq. 07.34.21.0003.01]{r121} after inserting the pdf of $w_{rd_i}$, we obtain $\Psi_{\textmd{1}}(\gamma_\textmd{S})$ as in (\ref{26}).
												
							\section{Proof of Theorem 3}
							On substituting $\Lambda_{D_i,2}$ as given by (\ref{eq24}) in (\ref{27}), we can evaluate $\mathcal{P}_2(\gamma_\textmd{S})$ as
							\begin{align}\label{35c}
								&	\mathcal{P}_2(\gamma_\textmd{S})=
								\textmd{Pr}\left[Y_{i}  > \frac{\sigma^2_{d_i}\eta_s\gamma_\textmd{S}Xw_{sr}^{-2}w^{\nu_{rd_i}}_{rd}}{\mathcal{P}_\textmd{th}\left(\mathcal{A}Xw_{sr}^{-2}-\mathcal{B}\right)},X> \frac{\mathcal{P}_
									\textmd{th}w_{sr}^{2}}{\eta_s} \right]. 
							\end{align}
							We hereby observe that the condition $\gamma_{\textmd{S}}\geq\mu^\prime$ results in $\mathcal{P}_2(\gamma_\textmd{S})=0$. For $\gamma_{\textmd{S}}< \mu^\prime$ and $\mathcal{P}_\text{th}\geq\left(\frac{\mathcal{B}}{\mathcal{A}}\right){\eta_s}$, we determine $\mathcal{P}_2(\gamma_\textmd{S})= \Psi_2(\gamma_{\textmd{S}})$ as
							\begin{align}\label{69a}
								\Psi_{\textmd{2}}(\gamma_\textmd{S})
								&=\int_{w_{\min}}^{w_{\max}}f_{w_{sr}}(w)\int_{h_0}^{\sqrt{h^2_0+l^2}}f_{w_{rd_i}}(v)\int_{\frac{\mathcal{P}_\textmd{th}w^{2}}{\eta_s}}^{\infty}  f_X(x) \int_{\frac{\sigma^2_{d_i}\eta_s\gamma_\textmd{S}v^{\nu_{rd_i}} x w^{-2}}{\mathcal{P}_\textmd{th}\left(\mathcal{A}xw^{-2}-\mathcal{B}\right)}}^{\infty}f_{Y_i}(y)dydxdvdw.
							\end{align} 
							By following a procedure similar to that used to obtain (\ref{39i}) in Appendix B, we can simplify (\ref{69a}) as 
							\begin{align}\label{39bb}
								\Psi_{\textmd{2}}&(\gamma_\textmd{S})
								=\sum _{k =0}^{m_{sr}-1} \int_{w_{\min}}^{w_{\max}}f_{w_{sr}}(w) \int_{h_0}^{\sqrt{h^2_0+l^2}}f_{w_{rd_i}}(v)\int_{\frac{\mathcal{P}_\textmd{th}w^2}{\eta_s}}^{\infty} \sum_{n=0}^{m_{rd_i}-1}\frac{1}{n!}\left(\frac{m_{rd_i}\sigma^2_{d_i}\eta_s\gamma_\textmd{S}xw^{-2}v^{\nu_{rd_i}}}{\mathcal{P}_\textmd{th}\left(\mathcal{A}xw^{-2}-\mathcal{B}\right)}\right)^n \nonumber\\
								&\times \exp\left(-\frac{m_{rd_i}\sigma^2_{d_i}\eta_s\gamma_\textmd{S}xw^{-2}v^{\nu_{rd_i}}}{\mathcal{P}_\textmd{th}\left(\mathcal{A}xw^{-2}-\mathcal{B}\right)}\right)f_X(x)dxdvdw.
							\end{align}  
							We now make a variable change $z=\mathcal{A}xw^{-2}-\mathcal{B}$ and substitute the pdf of $X$ in (\ref{39bb}). Subsequently, on applying the binomial theorem and Taylor series expansion for the term $\exp\left(-\frac{m_{rd_i}\sigma^2_{d_i}\eta_s\gamma_\textmd{S}v^{\nu_{rd_i}}}{\mathcal{P}_\textmd{th}\mathcal{A}z}\right)$, we get 
								\begin{align}\label{39b}
									&\Psi_{\textmd{2}}(\gamma_\textmd{S})
									=\alpha _{sr} \sum _{k =0}^{m_{sr}-1}   \zeta\left ({k }\right)\int_{w_{\min}}^{w_{\max}}f_{w_{sr}}(w)\int_{h_0}^{\sqrt{h^2_0+l^2}}f_{w_{rd_i}}(v) 
									\left( \frac{w^2}{\mathcal{A}} \right)^{k+1}\sum_{n=0}^{m_{rd_i}-1}\sum_{k_1=0}^{k+n}\binom{k+n}{k_1}\nonumber\\
									&\times\sum_{k_2=0}^{\infty }\frac{(-1)^{k_2}}{k_2!}\exp\left(\frac{-\bar{\beta}_{sr}\mathcal{B}w^2}{\mathcal{A}}\right)
									\left(\frac{m_{rd_i}\sigma^2_{d_i}\eta_s\gamma_\textmd{S}v^{\nu_{rd_i}}}{\mathcal{P}_\textmd{th}\mathcal{A}}\right)^{n+k_2} \frac{\mathcal{B}^{k+n-k_1+k_2}}{n!}\exp\left(-\frac{m_{rd_i}\sigma^2_{d_i}\eta_s\gamma_\textmd{S}v^{\nu_{rd_i}}}{\mathcal{P}_\textmd{th}\mathcal{A}}\right)\nonumber\\
									&\times
									\int_{\mathcal{P}_{\mathcal{A}, \mathcal{B}}}^{\infty }z^{-n+k_1-k_2}\exp\left(-\frac{\bar{\beta}_{sr}w^2z}{\mathcal{A}}\right)dzdvdw.
								\end{align} 
							In (\ref{39b}), we first solve the integral w.r.t. $z$ using \cite[eq. 3.383.4]{grad} and then, evaluate the resulting expression w.r.t. $v$ using \cite[eq. 2.33.10]{grad} after inserting the pdf of $w_{rd_i}$ to reach $\Psi_{\textmd{2}}(\gamma_\textmd{S})$ as in (\ref{e40}). 
							
							Furthermore, for $\gamma_{\textmd{S}}< \mu^\prime$ and $\mathcal{P}_\text{th}<\left(\frac{\mathcal{B}}{\mathcal{A}}\right){\eta_s}$, we determine $\mathcal{P}_2(\gamma_\textmd{S})= \Psi_3(\gamma_{\textmd{S}})$ as
							\begin{align}\label{39a}
								\Psi_{\textmd{3}} (\gamma_\textmd{S})
								&=\int_{w_{\min}}^{w_{\max}}f_{w_{sr}}(w)\int_{h_0}^{\sqrt{h^2_0+l^2}}f_{w_{rd_i}}(v)\int_{\frac{\mathcal{B}w^{2}}{\mathcal{A}}}^{\infty}  f_X(x) \int_{\frac{\sigma^2_{d_i}\eta_s\gamma_\textmd{S}v^{\nu_{rd_i}} x w^{-2}}{\mathcal{P}_\textmd{th}\left(\mathcal{A}xw^{-2}-\mathcal{B}\right)}}^{\infty}f_{Y_i}(y)dydxdvdw.
							\end{align}  
							By making a variable change $z=\mathcal{A}xw^{-2}-\mathcal{B}$ and substituting the pdf of $X$ in (\ref{39a}), we can obtain 
								\begin{align}\label{39d}
									\Psi_{\textmd{3}}(\gamma_\textmd{S})
									&=\alpha _{sr} \sum _{k =0}^{m_{sr}-1}  \zeta\left ({k }\right) \int_{w_{\min}}^{w_{\max}}f_{w_{sr}}(w)\int_{h_0}^{\sqrt{h^2_0+l^2}}f_{w_{rd_i}}(v)\exp\left( -\frac{\mathcal{B}\bar{\beta}_{sr}w^2}{\mathcal{A}}-\frac{m_{rd_i}\sigma^2_{d_i}\eta_s\gamma_\textmd{S}v^{\nu_{rd_i}}}{\mathcal{P}_\textmd{th}\mathcal{A}} \right)\nonumber\\
									&\times\sum_{n=0}^{m_{rd_i}-1}\sum_{k_{1}=0}^{k+n}\binom{k+n}{k_{1}}\frac{\mathcal{B}^{k+n-k_1}}{n!}\left( \frac{w^2}{\mathcal{A}}\right)^{k+1}\left(\frac{m_{rd_i}\sigma^2_{d_i}\eta_s\gamma_\textmd{S}v^{\nu_{rd_i}}}{\mathcal{P}_\textmd{th}\mathcal{A}}   \right)^{n}\nonumber\\
									&\times \int_{0}^{\infty }z^{k_1-n}\exp\left( -\frac{\bar{\beta}_{sr}w^2z}{\mathcal{A}}-\frac{m_{rd_i}\sigma^2_{d_i}\eta_s\gamma_\textmd{S}v^{\nu_{rd_i}}\mathcal{B}}{\mathcal{P}_\textmd{th}\mathcal{A}z} \right)dzdvdw.
								\end{align}
							Further, on solving the integral w.r.t. $x$ in (\ref{39d}) using \cite[eq. 3.471.9]{grad}, one can reach  
								\begin{align}\label{eq39ee}
									\Psi_{\textmd{3}}(\gamma_\textmd{S})
								&	=2\alpha _{sr} \sum _{k =0}^{m_{sr}-1}  \zeta\left ({k }\right)\sum_{n=0}^{m_{rd_i}-1}\frac{1}{n!} \sum_{k_{1}=0}^{k+n}\binom{k+n}{k_{1}} \int_{w_{\min}}^{w_{\max}}f_{w_{sr}}(w)\int_{h_0}^{\sqrt{h^2_0+l^2}}f_{w_{rd_i}}(v)\nonumber\\
									&\times\bar{\beta}_{sr}^{\frac{n-k_1-1}{2}}\frac{\left(\mathcal{B}w^2\right)^{\frac{2k+n-k_1+1}{2}}}{\mathcal{A}^{k+n+1}}\exp\left( -\frac{\mathcal{B}\bar{\beta}_{sr}w^2}{\mathcal{A}}-\frac{m_{rd_i}\sigma^2_{d_i}\eta_s\gamma_\textmd{S}v^{\nu_{rd_i}}}{\mathcal{P}_\textmd{th}\mathcal{A}}   \right)\nonumber\\
									&\times\left(\frac{m_{rd_i}\sigma^2_{d_i}\eta_s\gamma_\textmd{S}v^{\nu_{rd_i}}}{\mathcal{P}_\textmd{th}}  \right)^{\frac{n+k_1+1}{2}}\mathcal{K}_{k_1-n+1}\left( \sqrt{\frac{4m_{rd_i}\sigma^2_{d_i}\eta_s\gamma_\textmd{S}v^{\nu_{rd_i}}\mathcal{B}\bar{\beta}_{sr}w^2}{\mathcal{A}^{2}\mathcal{P}_\textmd{th}} } \right)dvdw.
								\end{align}
							The solution of (\ref{eq39ee}) in its current form is quite tedious. Therefore, we make use of the identity
							$\mathcal{K}_{\upsilon}(x)=\frac{1}{2}G^{2,0}_{0,2}
							\bigg[\frac{x^2}{4} \bigg\vert{{-}\atop{{\upsilon/2,-\upsilon/2}}}\bigg]$
							and the Taylor series expansion for the term $\exp\left(-\frac{m_{rd_i}\sigma^2_{d_i}\eta_s\gamma_\textmd{S}v^{\nu_{rd_i}}}{\mathcal{P}_\textmd{th}\mathcal{A}}\right)$ in (\ref{eq39ee}).
							Eventually, we evaluate the resulting expression w.r.t. $v$ using \cite[eq. 2.33]{grad} after inserting the pdf of $w_{rd_i}$ to obtain $\Psi_{\textmd{3}}(\gamma_\textmd{S})$ as in (\ref{e41}).		
							\section{Proof of Theorem 4}
							On substituting $\Lambda^\text{im-IC}_{T,1}$ as given by (\ref{21}) in (\ref{eq32}), we can evaluate $\mathcal{P}_3(\gamma_\textmd{A})$ as
							\begin{align}\label{25}
								\mathcal{P}_3(\gamma_\textmd{A})&=\textmd{Pr}\left[Z  > \frac{\sigma^2_{t}\gamma_{A}w^{\nu_{rt}}_{rt}}{\mathcal{C}Xw_{sr}^{-2}-\mathcal{D}},X\leq \frac{\mathcal{P}_\textmd{th}w_{sr}^{2}}{\eta_s} \right], 
							\end{align}
							where  $Z\triangleq\left| g_{rt} \right|^{2}$. For $\gamma_{\textmd{A}}< 1/\mu^\prime \mbox{ and } {\mathcal{P}_\text{th}}\geq\left(\frac{\mathcal{D}}{\mathcal{C}}\right){\eta_s}$, we determine $\mathcal{P}_3(\gamma_\textmd{A})=\widetilde{\Psi}_1(\gamma_\textmd{A})$ as 
							\begin{align}\label{e70}
									\widetilde{\Psi}_1(\gamma_\textmd{A})&={J}_1\left(h_1, {h_1}/{\cos\upphi}\right)+{J}_1\left({h_1}/{\cos\upphi}, h_2\right)+{J}_1\left(h_2, {h_2}/{\cos\upphi}\right),
								\end{align}
								where
								\begin{align}\label{38}
									{J}_1&\left(a,b\right)
									=\int_{a}^{b}f_{w_{rt}}(u)\int_{w_{\min}}^{w_{\max}}f_{w_{sr}}(w)\int_{\frac{\mathcal{D}w^{2}}{\mathcal{C}}}^{\frac{\mathcal{P}_\textmd{th}w^{2}}{\eta_s}}f_X(x)\int_{\frac{\sigma^2_{t}\gamma_{A}u^{\nu_{rt}}}{\mathcal{C}Xw^{-2}-\mathcal{D}}}^{\infty }f_{Z}(z)dzdxdwdu. 
								\end{align}  
								It is worth mentioning that the term ${J}_1\left(a,b\right)$ in (\ref{e70}) can be evaluated by following the similar procedure as given in Appendix B and hence, its detailed proof is omitted.	
								
								\section{Proof of Theorem 5}
								On substituting $\Lambda^\text{im-IC}_{T,2}$ as given by (\ref{22}) in (\ref{eq32}), we can evaluate $\mathcal{P}_4(\gamma_\textmd{A})$ as	
								\begin{align}\label{eq40}
									\mathcal{P}_4(\gamma_\textmd{A})&=
									\textmd{Pr}\left[Z  > \frac{\eta_{s}Xw_{sr}^{-2}\sigma^2_{t}\gamma_{A}w^{\nu_{rt}}_{rt}}{\mathcal{P}_\textmd{th}(\mathcal{C}Xw_{sr}^{-2}-\mathcal{D})},X> \frac{\mathcal{P}_\textmd{th}w_{sr}^{2}}{\eta_s} \right]. 
								\end{align}
								For $\gamma_{\textmd{A}}< 1/\mu^\prime \mbox{ and } {\mathcal{P}_\text{th}}\geq\left(\frac{\mathcal{D}}{\mathcal{C}}\right){\eta_s}$, we determine $\mathcal{P}_3(\gamma_\textmd{A})=\widetilde{\Psi}_2(\gamma_\textmd{A})$ as 
								\begin{align}\label{e72}
									\widetilde{\Psi}_2(\gamma_\textmd{A})&={J}_2\left(h_1, {h_1}/{\cos\upphi}\right)+{J}_2\left({h_1}/{\cos\upphi}, h_2\right)+{J}_2\left(h_2, {h_2}/{\cos\upphi}\right),
								\end{align}
								where
								\begin{align}\label{38}
									{J}_2&\left(a,b\right)
									=\int_{a}^{b}f_{w_{rt}}(u)\int_{w_{\min}}^{w_{\max}}f_{w_{sr}}(w)\int_{\frac{\mathcal{P}_\textmd{th}w^{2}}{\eta_s}}^{\infty}f_X(x)\int_{\frac{\eta_{s}xw^{-2}\sigma^2_{t}\gamma_{A}u^{\nu_{rt}}}{\mathcal{P}_\textmd{th}(\mathcal{C}xw^{-2}-\mathcal{D})}}^{\infty }
									f_{Z}(z)dzdxdudw. 
								\end{align}  
								
								Furthermore, for $\gamma_{\textmd{A}}< 1/\mu^\prime \mbox{ and } {\mathcal{P}_\text{th}}<\left(\frac{\mathcal{D}}{\mathcal{C}}\right){\eta_s}$, we determine $\mathcal{P}_4(\gamma_\textmd{A})=\widetilde{\Psi}_3(\gamma_\textmd{A})$ as
								\begin{align}\label{43}
									\widetilde{\Psi}_{3}(\gamma_\textmd{A})&=
									{J}_3\left(h_1, {h_1}/{\cos\upphi}\right)+{J}_3\left({h_1}/{\cos\upphi}, h_2\right)+{J}_3\left(h_2, {h_2}/{\cos\upphi}\right),
								\end{align} 
								where
								\begin{align}\label{e74}
									{J}_2&\left(a,b\right)
									=\int_{a}^{b}f_{w_{rt}}(u)\int_{w_{\min}}^{w_{\max}}f_{w_{sr}}(w)\int_{\frac{\mathcal{D}w^{2}}{\mathcal{C}}}^{\infty}f_X(x)\int_{\frac{\eta_{s}xw^{-2}\sigma^2_{t}\gamma_{A}u^{\nu_{rt}}}{\mathcal{P}_{th}(\mathcal{C}xw^{-2}-\mathcal{D})}}^{\infty }
									f_{Z}(z)dzdxdudw. 
								\end{align} 
								The solutions of $\widetilde{\Psi}_2(\gamma_\textmd{A})$ and $\widetilde{\Psi}_3(\gamma_\textmd{A})$ in (\ref{e72}) and (\ref{43}), respectively, follow the same procedure as in Appendix B and hence, the details are omitted. 
							\end{appendices}

						\end{document}